\documentclass[sigplan,twocolumn]{acmart}
\renewcommand\footnotetextcopyrightpermission[1]{}
\usepackage{tikz}
\usepackage{amsmath}

\usepackage{tikz}
\usepackage{amsmath}
\usepackage{mathtools}
\usepackage{amsthm}

\usepackage{hyperref}
\usepackage{booktabs}
\usepackage{makecell}
\usepackage{subfigure}
\usepackage[longend,ruled,vlined,linesnumbered]{algorithm2e}
\usepackage{multirow}
\usepackage{soul}
\usepackage{pifont}
\usepackage{xspace}
\usepackage{listings}
\newtheorem{example}{Example}
\newtheorem{definition}{Definition}

\newcommand{\eg}{{\it e.g.}, }
\newcommand{\ie}{{\it i.e.}, }

\newcommand{\TS}{Tiresias\xspace}
\newcommand{\SK}{\textsc{Tesserae}\xspace}
\newcommand{\ST}{\textsc{Tesserae-T}\xspace}

\newcommand{\SFair}{\textsc{Tesserae-FTF}\xspace}

\SetKw{Continue}{continue}
\SetKw{Break}{break}

\begin{document}

\date{}

\title{\SK: Scalable Placement Policies for Deep Learning Workloads}

\author{Song Bian}
\affiliation{%
  \institution{University of Wisconsin-Madison}
  \country{}
}
\email{songbian@cs.wisc.edu}
\author{Saurabh Agarwal}
\affiliation{%
  \institution{University of Wisconsin-Madison}
  \country{}
}
\email{agarwal@cs.wisc.edu}
\author{Md. Tareq Mahmood}
\affiliation{%
  \institution{University of Wisconsin-Madison}
  \country{}
}
\email{tareq@cs.wisc.edu}
\author{Shivaram Venkataraman}
\affiliation{%
  \institution{University of Wisconsin-Madison}
  \country{}
}
\email{shivaram@cs.wisc.edu}

\begin{abstract}

Training deep learning (DL) models has become a dominant workload in data-centers and improving resource utilization is a key goal of DL cluster schedulers. In order to do this, schedulers typically incorporate placement policies that govern where jobs are placed on the cluster. Existing placement policies are either designed as ad-hoc heuristics or incorporated as constraints within a complex optimization problem and thus either suffer from suboptimal performance or poor scalability. Our key insight is that many placement constraints can be formulated as graph matching problems and based on that we design novel placement policies for minimizing job migration overheads and job packing.
We integrate these policies into \SK and describe how our design leads to a scalable and effective GPU cluster scheduler. Our experimental results show that \SK improves average JCT by up to $1.62\times$ and the Makespan by up to $1.15\times$ compared with the existing schedulers.

\end{abstract}

\maketitle
\pagestyle{plain}

\section{Introduction}
\label{sec:introduction}


Training deep learning (DL) models~\cite{wang2018glue, lin2019commongen, cobbe2021training, wang2022super, russakovsky2015imagenet, sap2019socialiqa} has become a dominant workload in data-centers~\cite{jeon2019analysis,weng2022mlaas}. 
DL models are typically trained on large GPU clusters which are shared by several different jobs, and a scheduler is used to assign the resources to each DL training job.
Given the prevalence of DL jobs in data centers, several schedulers~\cite{qiao2021pollux, narayanan2020heterogeneity, zheng2023shockwave, mahajan2020themis, gu2019tiresias, xiao2018gandiva, xiao2020antman, hu2023lucid, jayaram2023sia} have been designed to tailor their policies to the unique characteristics of DL jobs.

The initial set of DL schedulers were designed to improve key scheduling metrics such as
job completion time (Tiresias~\cite{gu2019tiresias}) or finish-time fairness
(Themis~\cite{mahajan2020themis}). At a high level, these schedulers considered a set of queued jobs,
available cluster resources and various metrics (\eg attained service~\cite{gu2019tiresias}), and
selected a subset of jobs to execute on the cluster. While the scheduling algorithms used in these
works were successful at optimizing their respective target metrics, cluster operators found that GPU
utilization remained low~\cite{jeon2019analysis, weng2022mlaas, wesolowski2021datacenter} and
studies indicated that GPU utilization was closely linked to how and where jobs were
\emph{placed} in the cluster~\cite{xiao2018gandiva, mahajan2020themis}. 
Thus, in addition to scheduling policies, DL resource managers, 
also need to include placement policies that govern
where jobs are placed on the cluster, how they are migrated across scheduling rounds, 
the appropriate parallelization strategy used~\cite{zheng2022alpa} or which jobs are packed together~\cite{hu2023lucid}. 





In existing works, these placement policies are either implemented as ad-hoc heuristics, or incorporated
as a part of a larger joint optimization problem. 
For example, Tiresias~\cite{gu2019tiresias} uses a heuristic to classify which DL training jobs
require consolidated placement and Pollux~\cite{qiao2021pollux} uses a heuristic to minimize interference by ``ensuring at most one distributed job is allocated to each node''. On the other hand, Gavel~\cite{narayanan2020heterogeneity}
and POP~\cite{narayanan2021solving} jointly perform packing with scheduling by formulating the
packing constraints a part of their optimization problem.

Unfortunately, both approaches, adding as ad-hoc heuristics or incorporating placement
constraints within the optimization problem, has drawbacks. Introducing placement
constraints through ad-hoc heuristics often results in
suboptimal performance as these heuristics fail to capture the complex inter-dependencies among
scheduling decisions (Figure~\ref{fig:migration-exp-perf-limit}). Secondly, as the hardware and jobs
evolve,  these heuristics can become obsolete, and new heuristics might be needed, requiring manual
effort. For example, Tiresias uses a model's parameter count to determine placement, however, the
evolution in model architecture has led to this heuristic being obsolete~\cite{agarwal2023blox}.
Meanwhile, integrating placement constraints into the core scheduling optimization significantly
increases the number of variables while making the optimization problem more complex. 
This leads to poor scalability as the cluster sizes and number of jobs increase (Figure~\ref{fig:efficiency}).

These challenges highlight the need for a principled and scalable approach for integrating placement
policies in DL schedulers. Ideally, such an approach should exhibit the following properties. First,
the approach should provide a principled method for expressing a wide range of placement
constraints, offering better performance than existing heuristic methods and potentially matching the
performance of optimization-based solutions. Second, it should be adaptable to variations in hardware configurations, model architectures, and parallelization strategies. Third, it should be compatible with a broad range of scheduling policies to ensure wide applicability. Finally, the approach must be scalable to support modern GPU clusters with thousands of GPUs.


To overcome these challenges and attain these desirable properties, we introduce \SK. Our key
insight is that many placement constraints can be formulated as a \emph{graph matching} problem, which can be
efficiently solved using established algorithms such as the Hungarian
Algorithm~\cite{kuhn1955hungarian}. 
Precisely, we can build a graph $G=(V_1, V_2, E)$, where $V_1$ represents the set of jobs that are already placed on the cluster and $V_2$ represents the set of jobs that need to have a placement constraint applied.
For example, to minimize job migration, \SK  constructs a graph $G^{M} = (V^{M}_1, V^{M}_2, E^{M})$, where $V^{M}_1$ represents current placement plan and $V^{M}_2$ represents the placement plan following migration, the weight $w_{e^{M}}$ of edge $e^M = (x, y)$ quantifies the migration cost if the node $x$ from the original placement is reassigned to node $y$ in the new placement. 
With this formulation,
we show that we can formulate a number of placement policies, including
packing and migration using this approach and that such a graph-based formulation can be efficiently
solved even for large clusters. 
We incorporate our new placement policies and design \SK, a DL scheduler where placement policies take as input the jobs chosen by existing
scheduling policies such as Tiresias~\cite{gu2019tiresias}, Themis~\cite{mahajan2020themis} and come up with the final job placement. This design also allows DL schedulers to combine multiple possible placement policies.
We evaluate \SK using multiple workloads derived from prior schedulers~\cite{jeon2019analysis, narayanan2020heterogeneity,
zheng2023shockwave}. 
Our experiments first show that \SK's approach leads to significantly higher throughput than the
existing heuristic based placement policies improving 
JCT by $1.62\times$ and makespan by $1.15\times$. Next, we demonstrate that \SK can easily adapt to
hardware changes, such as varying GPU types, by showing that \SK adapts to altered GPU configuration
without requiring any additional tuning. Furthermore, we show that \SK is compatible with a range of
scheduling policies.
Finally, we evaluate the scalability of \SK and find that it remains efficient as the number of jobs
increases, taking less than 1.6 seconds for a cluster with 256 GPUs and 2048 jobs.

In summary, we make the following contributions: 
\begin{itemize}
    \item We first show that existing general-purpose linear programming-based schedulers such as Gavel~\cite{narayanan2020heterogeneity} and POP~\cite{narayanan2021solving} are not suitable at large scale due to the high overhead with the increased number of variables. (\S\ref{sec:challenges}) 

    
    \item We propose a novel graph-based formulation for the migration and packing algorithm, demonstrating that placement constraints can be resolved using a well-established weighted bipartite graph matching algorithm. (\S\ref{sec:optimization})
    \item Using physical cluster experiments and simulations we show that \SK can provide an improvement of up to $1.62\times$ on JCT and $1.15\times$ on Makespan while also improving the worst finish-time fairness ratio by $3.77\times$. (\S\ref{sec:evaluation})
\end{itemize}

\section{Background \& Motivation}
\label{sec:background_and_motivation}

In this section, we begin by providing an overview of deep learning scheduling in GPU clusters in \S\ref{sec:dl_scheduling}. Next, we highlight how scheduling policies are augmented by placement constraints to improve cluster utilization in \S\ref{sec:placement_policies}. Finally, we list the challenges with the way existing placement constraints have been defined in \S\ref{sec:challenges}.


\subsection{DL Scheduling}
\label{sec:dl_scheduling}

A multitude of studies~\cite{zheng2023shockwave, gu2019tiresias, hu2023lucid, xiao2018gandiva,
zhao2022multi, mahajan2020themis} have developed schedulers for DL workloads, targeting diverse
optimization goals. In DL scheduling, schedulers assign priorities to a set of jobs with the goal of
optimizing a specific performance metric. For example, Gandiva~\cite{xiao2018gandiva} focuses on
maximizing cluster utilization, whereas Tiresias~\cite{gu2019tiresias} is designed to reduce average
job completion time. 
Similarly, Pollux~\cite{qiao2021pollux} aims to optimize goodput—a metric combining system throughput and statistical efficiency.


To generalize a broad range of existing scheduling policies, Gavel~\cite{narayanan2020heterogeneity} introduces a linear programming framework that unifies optimization objectives, placement strategies, and packing policies within a single framework.
Gavel computes a priority score for each job based on the exact solution to an optimization problem
and the number of rounds of GPU allocation the job has received. 

\subsection{Placement Policies.}
\label{sec:placement_policies}

In addition to just scheduling, several  schedulers~\cite{xiao2018gandiva, gu2019tiresias,
mahajan2020themis} have highlighted the importance of placement constraints on the throughput
achieved by jobs. The placement constraints determine which GPUs in the cluster are used to run the
chosen jobs. We observe that placement constraints are handled primarily using two approaches: 

\paragraph{Heuristic based.}
Gandiva~\cite{xiao2018gandiva} depicted that distributed jobs prefer consolidated placements, \ie jobs requiring more than one GPU prefer that both GPUs exist on the same machine. Subsequently, Tiresias~\cite{gu2019tiresias} highlighted that certain jobs are more placement-sensitive than others, and cluster utilization can be improved by considering the properties of the individual job for performing placement. Furthermore, to mitigate low GPU utilization observed in production clusters~\cite{jeon2019analysis, weng2022mlaas}, several schedulers~\cite{xiao2018gandiva, hu2023lucid, narayanan2020heterogeneity} adopt GPU sharing techniques to better utilize available resources. When sharing GPUs, multiple jobs are concurrently run on the same GPUs.

\paragraph{Optimization based.}
Gavel~\cite{narayanan2020heterogeneity} demonstrates that specific placement constraints can be formulated as part of an optimization problem. Specifically, Gavel~\cite{narayanan2020heterogeneity} introduces packing multiple GPU jobs as a throughput aware optimization problem. 

However, both approaches— heuristic based algorithms and optimization based methods for handling placement constraints—pose certain challenges. In the following section, we discuss these challenges in detail.

\subsection{Challenges}
\label{sec:challenges}
We highlight the challenges associated with different type of placement constraints. 

\begin{figure}[!t]
    \centering
    \includegraphics[width=0.9\linewidth]{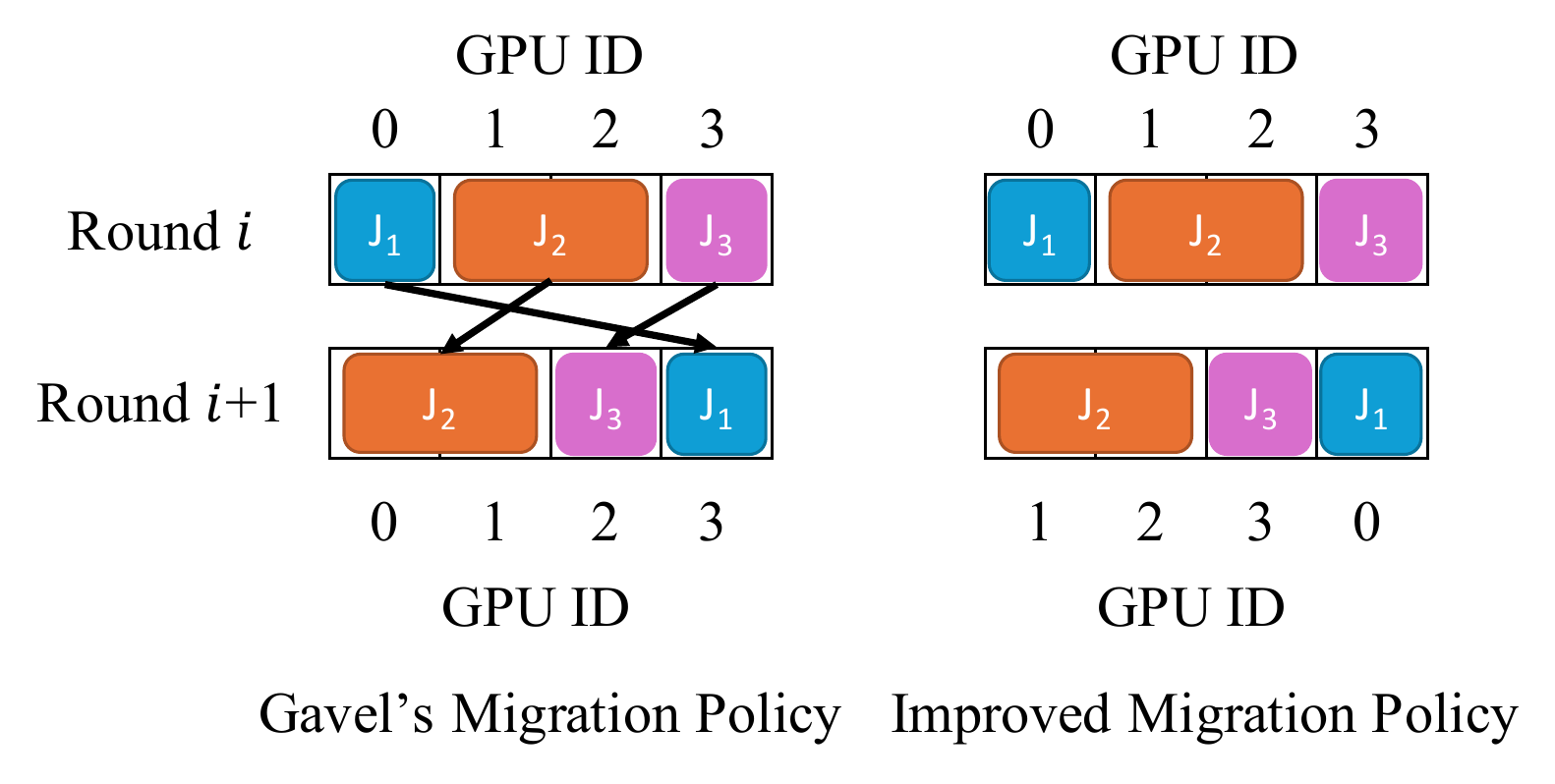}
    \vspace{-10pt}
    \caption{\small{\textbf{Performance Limitations:} In the left figure, Gavel’s policy migrates three jobs between two nearby plans. However, the right figure shows that we can avoid this overhead and improve throughput by remapping GPU ID. }}
    \label{fig:migration-exp-perf-limit}
    \vspace{-10pt}
\end{figure}

\paragraph{Performance Limitations}
We observe that heuristic-based placement approach often fail to maximize the possible throughput.
For example, Gavel's migration policy states that job migration is unnecessary if a job uses the same GPU in two consecutive placement rounds; otherwise, migration is required. However, in Figure~\ref{fig:migration-exp-perf-limit}, we observe that Gavel's migration policy~\cite{narayanan2020heterogeneity} results in three job migrations, whereas the optimal solution requires none. 


\paragraph{Poor Adaptability}
The deep learning ecosystem is rapidly evolving across the hardware stack, software stack, and
workloads. On the hardware side, GPUs have undergone significant advancements, resulting in varying
compute characteristics. For instance, newer GPUs support Multi-Instance GPU (MiG)~\cite{mig},
enabling fine-grained resource sharing. Meanwhile the software stack has also evolved, previously
distributed jobs used to use a parameter server setup to exchange
parameters~\cite{li2014communication}. Lately, the jobs have been using decentralized collective
communication call like all-reduce~\cite{sergeev2018horovod}. These software changes have led older
heuristics obsolete and require users to come up with new heuristics, requiring constant manual
intervention. Furthermore, as workloads transition between MLPs~\cite{naumov2019deep, agarwal2023bagpipe}, CNNs~\cite{krizhevsky2012imagenet}, transformers~\cite{vaswani2017attention}, and MoEs~\cite{shazeer2017outrageously}, users need to develop heuristics that consider their distinct compute and communication demands.

\begin{figure}[!t]
    \centering
    \includegraphics[width=0.8\linewidth]{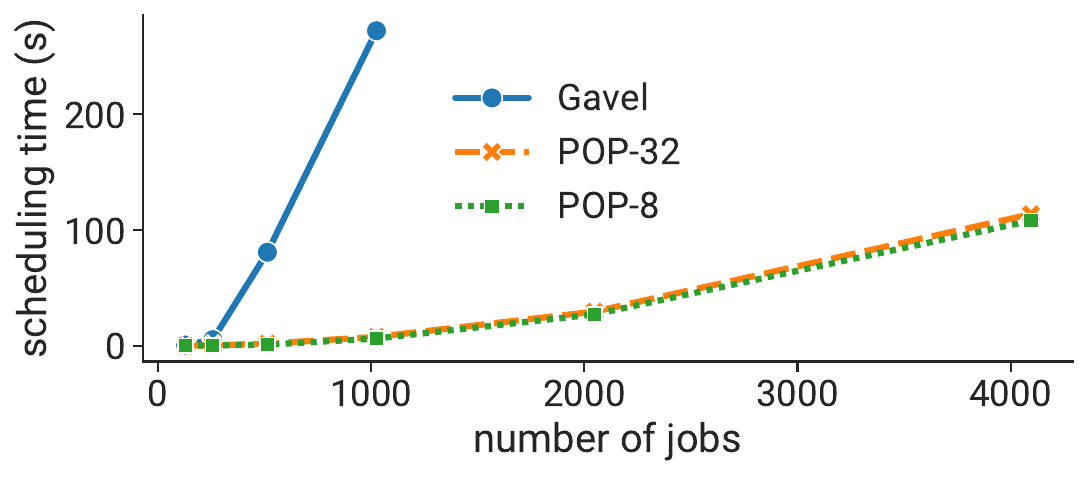}
    \vspace{-10pt}
    \caption{\small{\textbf{Overhead of Schedulers:} Decision-making time of each scheduler under varying numbers of active jobs in a 256-GPU cluster. The workload consists of jobs running ResNet-50, VGG-19, DCGAN, and PointNet, each with varying GPU requirements. The results indicate that both Gavel and POP exhibit limited scalability as the number of active jobs increases.
    }}
    \vspace{-10pt}
    \label{fig:efficiency}
\end{figure}

\paragraph{Limited Scalability.} Gavel's scalability is limited by the computational overhead of solving linear programs~\cite{cohen2021solving}. To overcome this limitation, POP~\cite{narayanan2021solving} is proposed as a scalable alternative to Gavel~\cite{narayanan2020heterogeneity}. However, as the number of GPUs and active jobs grows~\cite{hu2024characterization}, scheduler efficiency becomes a significant concern. As shown in Figure~\ref{fig:efficiency}, we fix the cluster size and vary the number of active jobs to evaluate the decision-making time of each scheduler. We observe that POP also faces challenges in scaling efficiently with an increasing number of active jobs—a limitation similarly observed in~\cite{kumar2024optimizing}. As a result, designing a truly scalable policy remains an open research challenge. 


\subsection{Goals}
These challenges highlight the need for a framework that enables users to specify placement constraints. In the following paragraphs, we describe the desirable characteristics of such an framework.

\paragraph{Improved Performance.}
A primary objective of the placement framework should be to enhance hardware utilization. Additionally, it should outperform widely adopted heuristics (\eg, Tiresias~\cite{gu2019tiresias}) and match the performance of linear programming-based solvers (\eg, Gavel~\cite{narayanan2020heterogeneity}). 


\paragraph{Adaptability.} The framework should be adaptable to the constant changes in the deep learning stack. It should automatically adjust placement constraints as the underlying hardware~\cite{V100, A100} and job types evolve~\cite{hu2024characterization}, without requiring manual intervention.

\paragraph{Compatibility}
The framework should also be compatible with existing scheduling policies, including Tiresias~\cite{gu2019tiresias}, and Themis~\cite{mahajan2020themis}. Users should be able to apply their chosen scheduler and use the framework solely for placement decisions, enabling wide adoption.

\paragraph{Scalability}
As deep learning clusters grow in size and handle an increasing number of jobs~\cite{weng2022mlaas, hu2024characterization}, the proposed framework demonstrates the ability to efficiently manage large-scale cluster scheduling.

\begin{figure}[!t]
\centering
\subfigure[Warmup and Checkpoint Time]{
\includegraphics[width=0.47\linewidth]{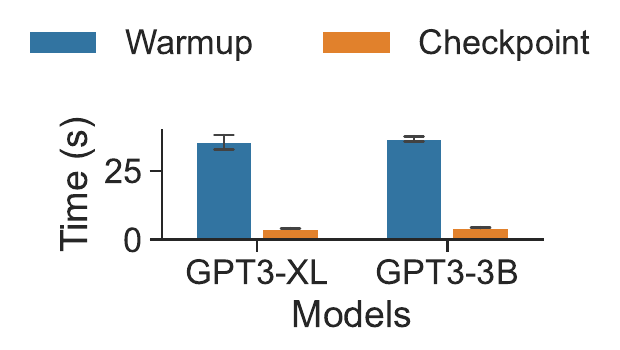}
\label{fig:overhead-warmup}
}
\subfigure[Number of Job Migrations]{  
\includegraphics[width=0.45\linewidth]{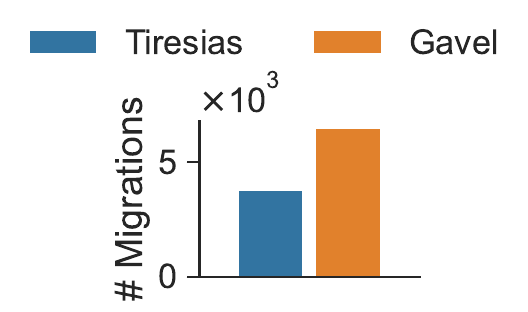} 
\label{fig:overhead-number}
}
\vspace{-10pt}
\caption{\small{\textbf{Migration Overhead:} The warmup time is the duration from entering the command to the start of the first iteration. Additionally, the checkpoint overhead represents the total time spent on loading and saving the checkpoint. Further, we evaluate the number of job migrations of Tiresias and Gavel.}}
\vspace{-10pt}
\label{fig:overhead}
\end{figure}

\begin{figure}[!t]
    \centering
    \includegraphics[width=\linewidth]{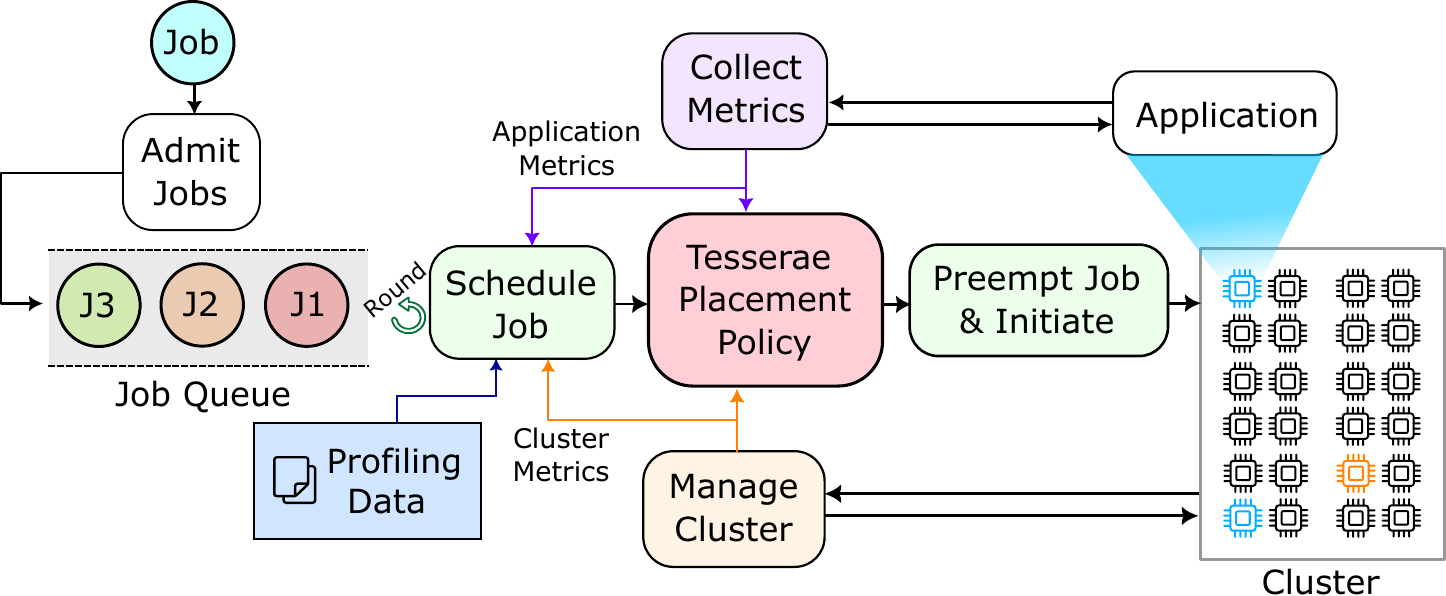}
    \vspace{-10pt}
    \caption{\small{\textbf{Overview of \SK system architecture:} all components of the system and their interactions. We design \SK as the placement policy for the whole system.}}
    \label{fig:shaka-workflow}
    \vspace{-10pt}
\end{figure}

\section{\SK}
\label{sec:shaka}
To tackle the challenges outlined in \S\ref{sec:background_and_motivation}, we present \SK. We first present an overview of our approach which involves separating scheduling policies from placement policies in \S\ref{sec:decompose}. Following that, in \S\ref{sec:goal_sk}, we present an overview of how scheduling in \SK works.



\subsection{Decomposing Scheduling and Placement}
\label{sec:decompose}

Existing DL schedulers typically take one of two approaches.  
First, the approach is the one taken by 
linear programming-based frameworks (e.g., Gavel~\cite{narayanan2020heterogeneity}).  These frameworks 
formulate the scheduling and placement policy as a single optimization problem. 
Formulating such an optimization problem enables capturing multiple objectives, such as minimizing job completion time (JCT) and improving GPU utilization through job packing. 
This approach is effective in making high-quality decisions, but it limits scalability due to high computational complexity.
The second approach is to treat scheduling and placement policies as distinct modules~\cite{gu2019tiresias, xiao2018gandiva, peng2018optimus, qiao2021pollux,agarwal2023blox}. 

In this work, we use the latter approach and view DL schedulers as a composition of different policies, such as a scheduling policy, consolidation policy, packing policy, etc. The benefit of this approach is that each policy can be independently designed and then composed together to build a scheduler (Figure~\ref{fig:shaka-workflow}). This approach aligns well with our design goals. 
First, the independent design of each policy ensures adaptability to changes in the deep learning stack. Second, it supports compatibility, allowing different scheduling objectives to be integrated with various placement policies. Moreover, if each policy is efficient and scalable, the overall scheduler inherits these properties.
Our evaluation (\S\ref{sec:evaluation}) also shows that our approach of treating scheduling and placement as disjoint policies does not adversely impact the quality of scheduling.  

\subsection{Overview}
\label{sec:goal_sk}
\label{sec:sys_overview}



Figure~\ref{fig:shaka-workflow} shows a high-level design overview of \SK . 
We next discuss a concrete example of how the \SK scheduler can apply policies such as migration-awareness and packing (Listing~\ref{alg:framework}).

\begin{figure}[!t]
    \centering
    \includegraphics[width=0.6\linewidth]{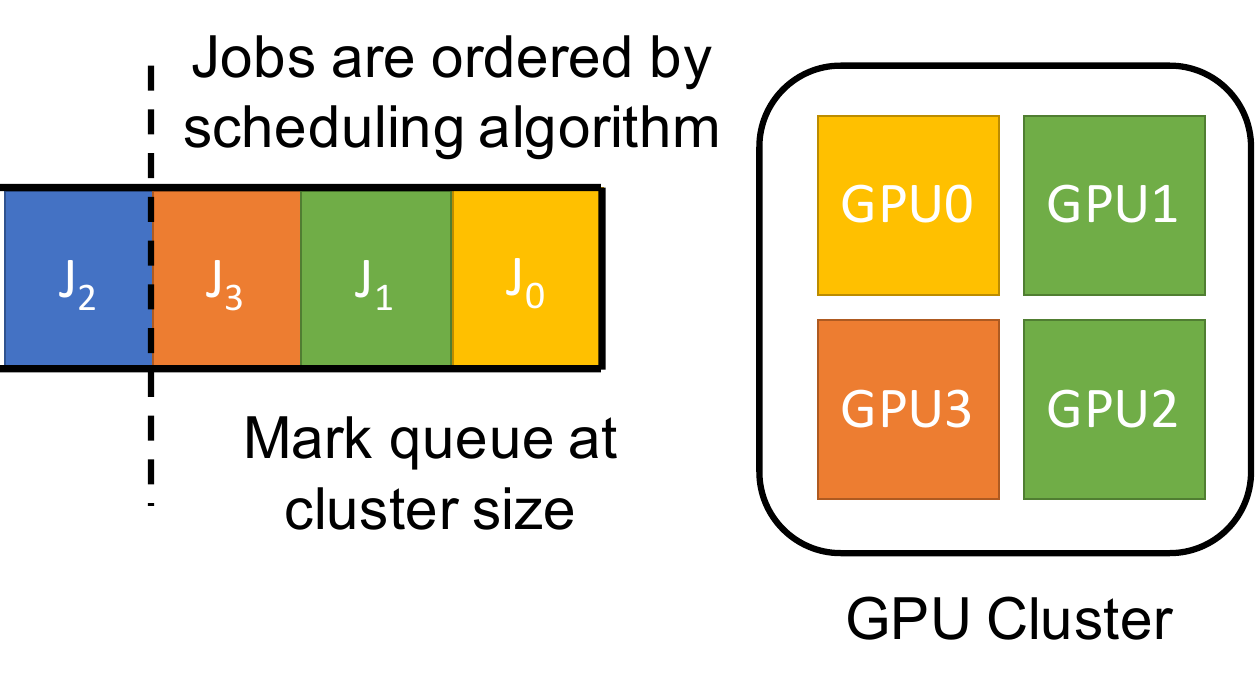}
    \vspace{-10pt}
    \caption{\small{\textbf{Allocation without Packing:} This example demonstrates how to allocate as many jobs as possible to a GPU cluster without GPU sharing. }}
    \label{fig:place-full}
    \vspace{-10pt}
\end{figure}

Scheduling in DL clusters typically happens in rounds~\cite{narayanan2020heterogeneity, mohan2022looking} (rounds are typically a few minutes). At the end of each round, the cluster scheduler is invoked to determine the set of jobs that should run for the next round.
In every round, the job scheduling policy sorts active jobs by their priority. The priority for jobs is determined based on the underlying scheduling policy, such as arrival time for FIFO or LAS for Tiresias~\cite{gu2019tiresias}, etc. 
Given the list of sorted jobs, a placement policy first places as many jobs as possible on the GPU cluster without packing, as depicted in Figure~\ref{fig:place-full}. 
This is because we would like to first place as many high-priority jobs on the cluster as we can to ensure scheduling priorities are met (lines 5-12 of Listing~\ref{alg:framework}). 
However, we note that placement (in line 8) can fail if there are fewer available GPUs than the number needed by the jobs or some placement constraint that cannot be satisfied, \eg a multi-GPU job cannot find a consolidated placement. After following these steps, given the significant migration overheads, we use our novel migration algorithm to minimizes the number of migrations between successive rounds.
Additionally, if GPU sharing is enabled, we use a packing policy to determine which jobs from \textsf{pending\_jobs} should be packed with already \textsf{placed\_jobs} on the GPU cluster, before determining the job migration strategy.

Next, we present the design of an efficient migration and packing policy that can operate alongside any scheduling policy.

\renewcommand{\algorithmcfname}{Listing}
\begin{algorithm}[t]
    \caption{\SK Framework}\label{alg:framework}
    \textsf{placed\_jobs} $\leftarrow []$ and \textsf{pending\_jobs} $\leftarrow []$\\
    \textsf{active\_jobs} $\leftarrow$ all submitted and unfinished jobs \\
    Sort \textsf{active\_jobs} based on priority \\
    \textsf{num\_gpus\_remain} $\leftarrow$ number of total GPUs \\
    \While{\textsf{num\_gpus\_remain} $> 0$}{
        \textsf{j} $\leftarrow$ \textsf{active jobs with highest priority} \\
        Remove job \textsf{j} from \textsf{active\_jobs} \\
        \If{we fail to place job \textsf{j}}{
            \textsf{pending\_jobs}.append(job \textsf{j}) \\
            \Continue \\
        }
        \textsf{num\_gpus\_remain} -= GPUs required by job \textsf{j} \\
        \textsf{placed\_jobs}.append(job \textsf{j}) \\
    }
    \If{GPU sharing is enabled}{
        \textsf{M} $\leftarrow$ \textsf{Packing}(\textsf{placed\_jobs}, \textsf{pending\_jobs}) \\
        Packing the jobs in \textsf{pending\_jobs} with jobs in \textsf{placed\_jobs} based on $\textsf{M}$ \\
    }
    Determine job migration strategy and place jobs across GPUs in the cluster \\ 
\end{algorithm}
\renewcommand{\algorithmcfname}{Algorithm}



\section{Efficient Migration and Packing Policies}
\label{sec:migration_packing}


We next describe new placement policies that can be integrated with \SK's design described before. Our scalable and performant placement policies are based on the insight that many placement problems can be viewed as instances of graph matching problems. We first present a migration algorithm (\S\ref{sec:migration}) to reduce the number of migrations and then introduce a packing policy (\S\ref{sec:optimization}) which can maximize throughput. Finally, we present approaches for minimizing profiling overhead and discuss the properties of \SK framework in \S\ref{sec:properties}.


\begin{algorithm}[t]
    \caption{\textsf{Job Migration}}
    \label{alg:job-migration}
    \KwIn{\textsf{placement\_plan for round $i$: $P_i$}, \textsf{placement\_plan for round $i+1$: $P_{i+1}$}.}
    \KwOut{\textsf{Migration Plan} \textsf{M}}
    \textsf{C} $\leftarrow []$, \textsf{M} $\leftarrow []$ \\
    \textsf{Remove all jobs $j$ in $P_i$ and $P_{i+1}$, where job j satisfies} $j \in ((P_i \cup P_{i+1}) - (P_{i} \cap P_{i+1}))$ \\
    \ForEach{placement plan of node $k$, $P_{i,k}$ $\in$ \textsf{$P_i$}}{
        \ForEach{placement plan of node $\ell$, $P_{i+1,\ell} \in P_{i+1}$}{
            $C_{k,\ell}, M_{k, \ell} \leftarrow$ \textsf{Node-Level Matching}($P_{i,k}$, $P_{i+1, \ell}$)
        }
    }
    \textsf{match\_solution} $\leftarrow$ \textsf{Hungarian Algorithm}(\textsf{C}) \\
    \Return \textsf{M[match\_solution]}
\end{algorithm}

\begin{algorithm}[t]
    \caption{\textsf{Node-Level Matching}}
    \label{alg:node-level}
    \KwIn{\textsf{placement\_plan for round $i$ of node $k$: $P_{i,k}$}, \textsf{placement\_plan for round $i+1$ of node $\ell$: $P_{i+1, \ell}$}.}
    \KwOut{\textsf{Migration Cost $C_{k,\ell}$}, \textsf{Migration Plan $M_{k,\ell}$}}
    $C_{\text{sum}} \leftarrow 0$, $C \leftarrow [0]_{k_{\ell} \times k_{\ell}}$, where $k_{\ell}$ is the number of GPUs in the node. \\
    \ForEach{GPU \textsf{$u$} $\in$ \textsf{$P_{i,k}$}}{
        \ForEach{GPU \textsf{$v$} $\in$ \textsf{$P_{i+1,\ell}$}}{
            $JS_{u} \leftarrow$ Job sets on GPU $u$ and $JS_{v} \leftarrow$ Job sets on GPU $v$ \\
            \ForEach{job \textsf{$j$} $\in JS_u \cup JS_v$}{
                \If{job \textsf{$j$} $\in$ $((JS_u \cup JS_v) - (JS_u \cap JS_v))$}{
                    $C_{u,v} \leftarrow C_{u,v} + 1 / (2 \cdot \textsf{num\_GPUs($j$)})$ \\
                }
            }
        }
    }
    $C_{\text{sum}}$, $M_{k,\ell}$ $\leftarrow$ \textsf{Hungarian Algorithm}(\textsf{C}) \\
    \Return $C_{\text{sum}}$, \textsf{$M_{k,\ell}$}
\end{algorithm}

\subsection{Minimizing Migrations}
\label{sec:migration}



We first introduce a novel and efficient migration algorithm designed to decrease the number of job migrations between consecutive rounds. To begin with, we define job migration below:
\begin{definition}[Job Migration]
A job is migrated between consecutive rounds if it is present in both rounds and utilizes different sets of GPUs within the cluster.
\end{definition}
Note that if a job does not appear in both round $i$ and round $i+1$, it is not considered to have been migrated. Next, we present the problem we study in this section:
\begin{definition}[Job Migration Minimization]
Given two placement plans, $P_i$ from round $i$ and $P_{i+1}$ from round $i+1$, the job migration minimization problem aims to determine a migration strategy that reduces the number of job migrations between these rounds while still meeting the constraints of consolidated placement.
\end{definition}

\begin{figure*}[!t]
    \centering
    \includegraphics[width=0.8\linewidth]{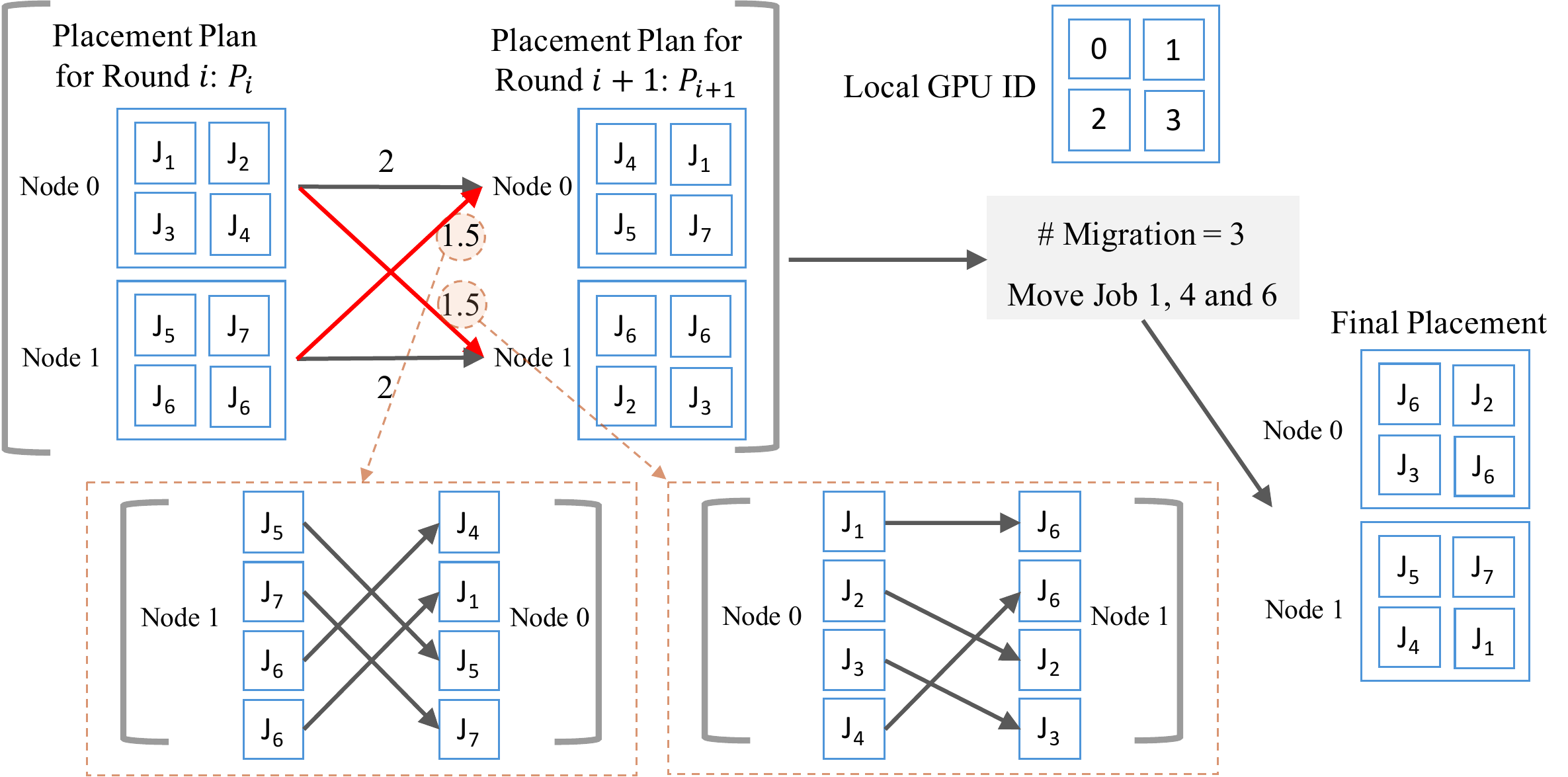}
    \vspace{-10pt}
    \caption{\small{\textbf{An Example of Migration Method:} Given two placement plans $P_i$ and $P_{i+1}$ from consecutive round $i$ and $i+1$, we show how to use Algorithm~\ref{alg:job-migration} and \ref{alg:node-level} to compute the migration plan and get the final placement plan in the end.}}
    \vspace{-10pt}
    \label{fig:migration_example}
\end{figure*}



We design a migration algorithm based on the following observation: say we have two placement plans from round $i$ and round $i+1$, as $\{(0, 1), (1, 2), (2, 2), (3, 4)\}$ and $\{(0, 4), (1, 1), (2, 2), (3, 2)\}$, respectively. In each tuple, the first number represents the GPU ID and the second number represents the job ID. We assume GPUs in the cluster are homogeneous. Although the two placement plans differ, job migrations are unnecessary as we can simply rename the GPU IDs through the following reassignments: $0 \rightarrow 1$, $1 \rightarrow 3$, and $3 \rightarrow 0$. Therefore, it is unnecessary to migrate any jobs. In view of this, we draw inspiration from established assignment problems and show our migration algorithm in Algorithm~\ref{alg:job-migration}. In particular, the algorithm first removes any jobs that are not present in both rounds concurrently (line 2 of Algorithm~\ref{alg:job-migration}). Next, the algorithm computes the migration cost for every pair of nodes, where each pair consists of one node $k$ from round $i$ and one node $\ell$ from round $i+1$ (lines 3-5 of Algorithm~\ref{alg:job-migration}). We use Algorithm~\ref{alg:node-level} to compute the migration cost between node pairs. Specifically, we first calculate the migration cost for each GPU pair across two given nodes (lines 2–7 of Algorithm~\ref{alg:node-level}). The cost is determined by the number of GPUs required by each job, as the migration cost is amortized across all processes in a multi-GPU job. Moreover, each move-in or move-out operation incurs a cost of 0.5 per job, which is why we multiply the cost by $1/2$ in line 7 of Algorithm~\ref{alg:node-level}. Finally, we apply the Hungarian algorithm~\cite{kuhn1955hungarian} to determine the optimal GPU-level migration between node pairs. 

After determining the migration cost for every pair of nodes, we can then apply the Hungarian algorithm~\cite{kuhn1955hungarian} again to derive a migration plan that minimizes the total number of migrations between the two placement plans (lines 6-7 of Algorithm~\ref{alg:job-migration}). We illustrate the complete process using the following example.



\begin{example}
\label{exp:migration}
We illustrate our migration strategy with an example in Figure~\ref{fig:migration_example}. Specifically, lines 3–5 of Algorithm~\ref{alg:job-migration} are used to compute the migration cost between each node pair. For instance, the migration cost between node 0 in round $i$ and node 1 in round $i+1$ is 1.5, as jobs 1 and 4 must be moved out and job 6 must be moved in. Each move-in or move-out operation incurs a cost of 0.5 per job. The corresponding migration plan between node 0 and node 1 is also shown in Figure~\ref{fig:migration_example}. Finally, using line 6 of Algorithm~\ref{alg:job-migration}, we determine the total number of migrations to be 3, along with the associated migration plan.

\end{example}




\paragraph{Running Time.} We assume that the cluster has $k_c$ nodes and each node contains $k_{\ell}$ GPUs. The Hungarian algorithm~\cite{kuhn1955hungarian} is widely used to address the assignment problem, and the time complexity of the Hungarian algorithm is $O(n^3)$, where $n$ is the number of tasks in the assignment problems. Therefore, the time complexity of Algorithm~\ref{alg:node-level} is $O(k_{\ell}^3)$ and the time complexity of Algorithm~\ref{alg:job-migration} is $O(k_c^2 k_{\ell}^3 + k_c^3)$.



\begin{figure*}[!t]
    \centering
    \includegraphics[width=0.8\linewidth]{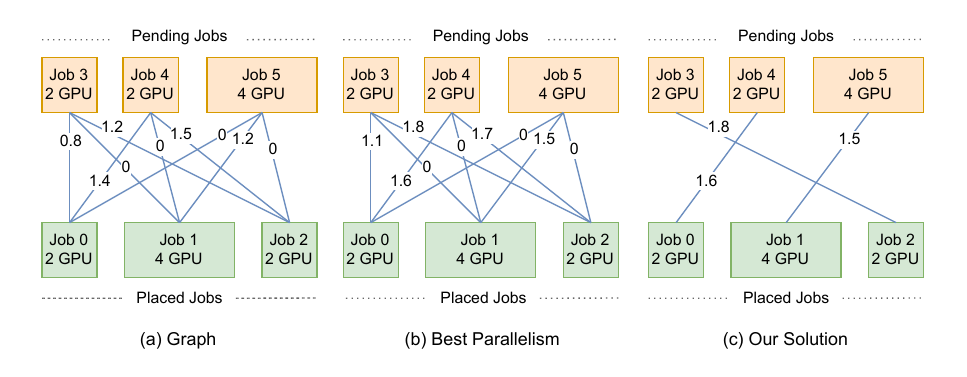}
    \vspace{-10pt}
    \caption{\small{\textbf{An Example of Job Packing:} Packing plans are developed by formulating them as weighted bipartite graph matching problems, where the weight of each edge represents the combined throughput of two jobs. We show the matching results obtained from our designed strategy.}}
    \vspace{-10pt}
    \label{fig:graph_example}
\end{figure*}

\begin{algorithm}[t]
    \caption{\textsf{Packing}}
    \label{alg:optimal}
    \KwIn{\textsf{placed\_jobs}, \textsf{pending\_jobs}, 
    \textsf{profile\_data}}
    \KwOut{Matching Results \textsf{M}}
    \textsf{G} $\leftarrow$ an empty graph \\ 
    \ForEach{job \textsf{j} $\in$ \textsf{placed\_jobs}}{
        \textsf{G.addNode(j)} \\
    }
    \ForEach{job \textsf{pj} $\in$ \textsf{pending\_jobs}}{
        \textsf{G.addNode(pj)} \\
        \ForEach{job \textsf{$j$} $\in$ \textsf{placed\_jobs}}{
            \If{\textsf{j} and \textsf{pj} require the same number of GPUs}{
                \textsf{w} $\leftarrow$ the sum of throughput of \textsf{pj} and \textsf{j} from \textsf{profile\_data} \\
                \textsf{G.addEdge(pj, j, w)} \\
            }
        }
    }
    \textsf{M} $\leftarrow$ computing maximum weighted bipartite graph matching with Hungarian Algorithm \\
    \Return \textsf{M}
\end{algorithm}

\subsection{Packing Jobs Efficiently}
\label{sec:optimization}

We next introduce a novel packing algorithm which uses a graph-based formulation to scale to a large number of jobs.

\paragraph{Profiling.} Our initial step in achieving optimized packing results is to profile and estimate the throughput of job packing. To normalize the throughput, we divide the packed throughput of jobs by their isolated throughput. For example, without packing, consider the throughput of PointNet to be 50 iterations per second, while GPT3-3B has a throughput of 2 iterations per second. With packing, say the throughput of PointNet drops to 15 iterations per second, while GPT3-3B drops to 1 iteration per second. Consequently, the normalized throughput of PointNet is 0.3 and GPT3-3B is 0.5. The combined or sum throughput in this case is 0.8. Our definition of normalized throughputs is similar to that used in prior work~\cite{narayanan2020heterogeneity} and in the following section, we will use normalized throughputs to formulate our graph-based problem.

\paragraph{Graph-based Problem Formulation.} We first reiterate the goal of the packing problem:  given a list of \textsf{placed\_jobs} and \textsf{pending\_jobs}, we wish to pack jobs such that we can maximize the total cluster throughput.



We observe that the job packing problem can be converted to maximum weighted bipartite graph matching problem. To be specific, we build a graph $G=(V_1, V_2, E)$, where $v_1 \in V_1$ represents a job in \textsf{placed\_jobs}, $v_2 \in V_2$ stands for a job in \textsf{pending\_jobs}, and $e = (u, v) \in E$ indicates that we can pack job $u$ and job $v$ on the same set of GPUs. In other words, job $u$ and job $v$ require the same number of GPUs. The weight $w_e$ of edge $e = (u, v)$ equals the combined throughput of job $u$ and job $v$, as determined by profiling the combined throughput of job $u$ and job $v$. Figure~\ref{fig:graph_example}(a) illustrates a graph constructed using profiling data. 





\begin{figure}[t]
\centering
\includegraphics[width=0.85\linewidth]{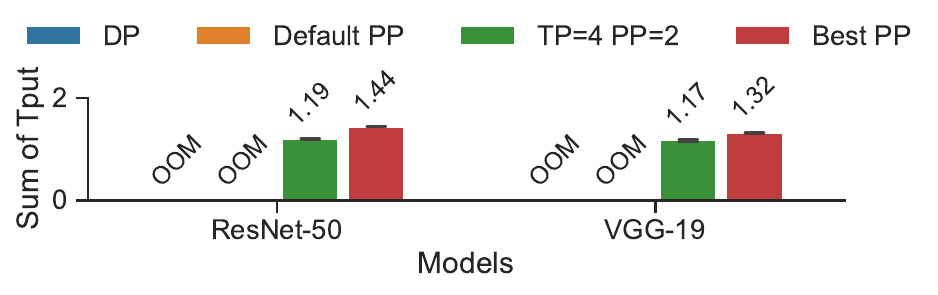}
\vspace{-10pt}
\caption{\small{\textbf{Throughput with packing:} We evaluate performance of training language models under different parallelization strategies with packing on 8 A100 GPUs. The Default PP is provided by Megatron-LM~\cite{shoeybi2019megatron} and the Best PP is picked from the candidate of possible PP strategies. The throughput is normalized by the best performance achieved in isolation. We observe that packing under certain scenarios can improve total throughput from the cluster}}
\label{fig:throughput}
\vspace{-10pt}
\end{figure}

\paragraph{Parallelism Strategy.} Here, we consider a new dimension introduced by 3D parallel training jobs: that of changing the parallelism strategy and studying how that affects packing. Prior work~\cite{zheng2022alpa, jia2019beyond, wang2019supporting} has studied the problem of choosing the best parallelization strategy to improve the throughput of a given LLM training job. We consider how this degree of freedom in choosing parallelization strategies can affect packing. We conduct a benchmarking study where we use model GPT3-3B and consider an 8 GPU setup. Figure~\ref{fig:throughput} shows the results of the above study. We observe that selecting an appropriate parallelism strategy not only prevents out-of-memory (OOM) issues-such as those observed when packing VGG-19 using the default pipeline-parallel (PP) strategy but also significantly boosts overall throughput. For instance, by adopting the parallelism strategy $\text{PP}=(3,3,3,4,4,5,5,5)$, which specifies the number of layers on each GPU, the sum of normalized throughput for ResNet-50 packed with GPT3-3B increases from $1.19$ to $1.44$. 

We can easily integrate the determination of the parallelization strategy into the previously described graph-based problem. If job $u$ is packed with job $v$, we have to modify the edge weight $w_e$ of edge $e=(u,v)$ in the constructed bipartite graph $G$, when optimizing the parallelism strategy of job $u$. In Figure~\ref{fig:graph_example}(b), we demonstrate the enhancement of an edge's weight through the choice of the best packing parallelism strategy. For example, selecting the best parallelism strategy for Job 1 can enhance the weight $w_{e'}$ of the edge between job $1$ and job $5$ from 1.2 to 1.5.


\paragraph{Solving Graph-based Problems.}
For each scheduling round, given $n$ active jobs and the number of GPUs in the cluster, the scheduling policy, which determines the priority of each job, creates \textsf{placed\_jobs} and \textsf{pending\_jobs}. Therefore, $V_1$ and $V_2$ in $G$ are fixed with each edge’s weight established based on offline profiling data. To solve such a bipartite graph matching problem, we use the classic Hungarian Algorithm~\cite{kuhn1955hungarian}. The time complexity of the Hungarian Algorithm is $O(n^3)$, where $n$ is the number of nodes in the graph. Solving the weighted bipartite graph matching problem yields a solution with each \textsf{placed\_job} is matched with at most one \textsf{pending\_job} and the algorithm ensures that overall sum of the weights of the chosen edges is maximized in the solution, thus yielding the maximum combined throughput from packing. Algorithm~\ref{alg:optimal} presents our packing algorithm, and its resulting solution for the example in Figure~\ref{fig:graph_example}(b) is illustrated in Figure~\ref{fig:graph_example}(c).

Furthermore, in Figure~\ref{fig:efficiency}, we compare the overhead associated with \SK against Gavel~\cite{narayanan2020heterogeneity} and POP~\cite{cohen2021solving}. We can see that the use of the Hungarian algorithm for packing decisions marginally increases the overhead already present in existing scheduling algorithms. We also see that \SK is more efficient than Gavel because \SK involves fewer variables. Although POP~\cite{narayanan2021solving} is designed to speed up the Gavel solver, it remains less efficient compared to \SK. Notably, even with 3000 active jobs, 
\SK is capable of making placement decisions within 1 second, which means that \SK can be deployed in large-scale cluster management systems.

\subsection{Discussion}

\paragraph{Minimizing Profiling Cost.}
\label{sec:profile_cost}
Although incorporating parallelism strategies into the maximum weighted bipartite graph matching problem is simple, profiling all such strategies offline is impractical. To reduce the profiling cost, we use the following two strategies:

For models trained by using only data parallelism, e.g. ResNet-50, VGG-19, we build the estimation model based on the following assumption: \textsf{If the model and GPU type are the same, the throughput of the 2-GPU job is double that of the 1-GPU job}~\cite{jayaram2023sia}. Therefore, for Job $J$, we first profile it on a single GPU to determine its throughput, denoted as $\textsf{tput}_{J}$. Then, we use the mathematical model to predict the throughput of Job $J$ over $N$ GPUs: $\textsf{tput}_{J}(N) = N \times \textsf{tput}_{J}$. 

Furthermore, if Job $J$ is packed with Job $K$, then we first profile them on a single GPU to determine their packed throughput, denoted as $\widehat{\textsf{tput}}_{J}$ and $\widehat{\textsf{tput}}_{K}$ respectively. Next, we estimate the throughput of Job $J$ and Job $K$ packed over $N$ GPUs as
$\widehat{\textsf{tput}}_{J}(N) = N \times \widehat{\textsf{tput}}_{J}$ and $\widehat{\textsf{tput}}_{K}(N) = N \times \widehat{\textsf{tput}}_{K}$.

Figure~\ref{fig:throughput} shows that when using an identical number of GPUs, selecting an optimal parallelism strategy can significantly increase the throughput of large language models. Consequently, the linear model, which consistently yields the same throughput for a given number of GPUs, is inadequate for models trained using 3D parallelism. To reduce profiling costs, we first profile large language models with randomly generated strategies. We then use Bayesian Optimization~\cite{snoek2012practical} to iteratively profile the model with subsequent parallelism strategies until the profiling budget is exhausted. The above strategy is similar to parameter tuning approaches developed in prior works~\cite{van2017automatic, kanellis2022llamatune}. We evaluate the effectiveness of our profiling optimizations in \S\ref{sec:ablation}.


\label{sec:properties}


\paragraph{Extensibility.} \SK is compatible with numerous established scheduling policies. For policies that do not account for packing, it suffices to modify the sorting priority at line 3 of Algorithm~\ref{alg:framework}. Moreover, \SK is equivalent to heuristic policies if we do not execute \textsf{Packing} function in Algorithm~\ref{alg:framework} since we use the same strategy as heuristic policies to order the active jobs. Additionally, we can also use Gavel to compute priority score to order the active jobs without considering GPU sharing. We show how \SK can work with LAS-based schedulers (e.g., Tiresias~\cite{gu2019tiresias}) and fairness-based schedulers  (e.g., Themis~\cite{mahajan2020themis}) in Section~\ref{sec:evaluation}.

\paragraph{Fairness.} Compared with isolated execution, the packing policy increases the total throughput of packed jobs, but it could reduce the throughput of each job. For jobs with high priority or strict deadlines, we can bypass the packing process by not creating edges between such a job and others when formulating the graph in Algorithm~\ref{alg:optimal}.


\paragraph{Consolidated Placement.} If the jobs from placement plan $P_{i}$ and $P_{i+1}$ in rounds $i$ and $i+1$ respectively are consolidated, Algorithm~\ref{alg:job-migration} will ensure that all jobs remain in a consolidated setting. Because Algorithm~\ref{alg:node-level} performs GPU matching at the node level, it ensures that the processes of a distributed job either remain on the same node or are collectively relocated to other nodes.

\section{Implementation}
\label{sec:implement}

\SK is built on Blox~\cite{agarwal2023blox} using Python in approximately 3000 lines of code for both real cluster mode and simulation mode. We have incorporated \SK into the Blox framework by developing it as a placement policy. We use Scipy~\cite{virtanen2020scipy} to generate the migration plan for each round outlined in \S\ref{sec:migration} and solve the weighted bipartite graph matching problem mentioned in \S\ref{sec:optimization}. Moreover, we employ gRPC~\cite{grpc} for message communication between schedulers and applications. Lastly, we use CUDA-MPS~\cite{mps} to run multiple jobs on the same GPU.

\paragraph{Profiling.} 
In this paper, we adopt a simple approach to collect profiling data by running all job combinations offline. \SK accumulates profiling data in an offline mode. It operates language models under various parallelism strategies, running each model listed in Table~\ref{tab:model} and each possible model combination for three minutes to measure their respective throughput. However, collecting all profiling data offline is impractical in real applications. We use strategies mentioned in \S\ref{sec:profile_cost} to reduce profiling costs and we evaluate these strategies in \S\ref{sec:ablation}.




\paragraph{Schedulers.} \SK  like  prior round based schedulers~\cite{narayanan2020heterogeneity, xiao2020antman}  make scheduling decisions every six minutes, 
taking the following steps: 
For all active jobs, it develops packing and placement plans based on the offline profiling data and the given scheduling method. The scheduler, after formulating the packing and placement plans, will notify all nodes to stop current jobs and start new ones. Note that \SK only preempts the job after the job finishes the current iteration. Then, the scheduler pauses for six minutes to gather updated metrics, such as service attained and training progress. Following that, it makes decisions for the next round.


\paragraph{Applications.} All applications are implemented using PyTorch~\cite{paszke2019pytorch}.  For all models listed in Table~\ref{tab:model}, \SK classifies the model into two groups depending on the presence of Transformer layers~\cite{vaswani2017attention}, leading to models being divided into two groups: (1) ResNet-50, VGG-19, DCGAN, and PointNet; (2) GPT3-Medium, GPT3-XL, and GPT3-3B. For the first group, distributed training is conducted using PyTorch DDP. In contrast, for the second group, 3D parallelism training is implemented using Megatron-LM~\cite{shoeybi2019megatron}. The parallelism strategy can be adjusted before launching jobs on the cluster. 
Furthermore, \SK imposes a limit of two models running simultaneously on each GPU because packing more than two jobs typically does not provide additional benefits~\cite{narayanan2020heterogeneity, hu2023lucid}.

\begin{table}[!t]
\centering
\caption{Models used in the evaluation. $\clubsuit$: Image Classification, $\diamondsuit$: Image-to-Image Translation, $\heartsuit$: 3D Point Cloud Classification, $\spadesuit$: Language Modeling.}
\vspace{-10pt}
\begin{tabular}{c c c c}
\toprule
Model & Task & Dataset & Batch Size  \\ 
\midrule
ResNet-50~\cite{he2016deep} & $\clubsuit$ & ImageNet~\cite{deng2009imagenet} & 32-256  \\
VGG-19~\cite{simonyan2014very} & $\clubsuit$ & ImageNet~\cite{deng2009imagenet} & 16-128 \\
DGCAN~\cite{radford2015unsupervised} & $\diamondsuit$ & LSUN~\cite{yu2015lsun} & 128-1024 \\ 
PointNet~\cite{qi2017pointnet} & $\heartsuit$ & ShapeNet~\cite{chang2015shapenet} & 32-256  \\
GPT3-Medium~\cite{brown2020language} & $\spadesuit$ & Wikipedia~\cite{devlin2018bert} & 512 \\
GPT3-XL~\cite{brown2020language} & $\spadesuit$ & Wikipedia~\cite{devlin2018bert} & 512 \\
GPT3-3B~\cite{brown2020language} & $\spadesuit$ & Wikipedia~\cite{devlin2018bert} & 512 \\
\bottomrule
\end{tabular}
\vspace{-10pt}
\label{tab:model}
\end{table}

\section{Evaluation}
\label{sec:evaluation}

In this section, we evaluate \SK and focus on the following aspects - (i) The effectiveness of \SK on a real cluster; (ii) The impact of each of our performance optimizations on \SK's efficiency; (iii) The adaptability and compatibility of \SK; (iv) The scalability of \SK.

\subsection{Experimental Setup} 
\label{sec:exp_setup}

\paragraph{Testbed.} We conduct cluster experiments using 32 GPUs across 8 nodes on NERSC Perlmutter~\cite{nersc}. 
Each node has four 40 GB NVIDIA A100 (Ampere) GPUs, 256 GB DDR4 DRAM, and a single AMD EPYC 7763 (Milan) CPU. 
We also run simulation experiments on a Cloudlab server~\cite{duplyakin2019design}. 
The server contains 188 GB memory and two Intel Xeon Silver 4114 10-core CPUs at 2.20 GHz.

\paragraph{Traces.} 
To evaluate the schedulers we use traces introduced by two prior works Shockwave~\cite{zheng2023shockwave} and Gavel~\cite{narayanan2020heterogeneity}. Table~\ref{tab:model} lists the workload details including models, dataset, and batch size. 
Our default trace is similar to Shockwave~\cite{zheng2023shockwave}, we follow the same setting as Shockwave by setting the probability of generating Small, Medium, Large, and Extra Large jobs to be 0.72, 0.2, 0.05, and 0.03. We also set the probability of generating 1-GPU, 2-GPU, 4-GPU, and 8-GPU jobs to be 0.6, 0.3, 0.09, and 0.01 to align with those used in Shockwave. The job arrival rate is configured at 80 jobs per hour, consistent with the settings used in prior work. We show how \SK's benefits change across workloads by using an additional trace similar to the one used in Gavel~\cite{narayanan2020heterogeneity} in \S\ref{sec:ablation}.
By default, we use a 120 jobs trace for physical experiments and a 900 jobs trace for simulated experiments, both with a job arrival rate of 80 jobs per hour. 

\paragraph{Baselines.} We compare \SK with \emph{four} different baselines (i) Tiresias~\cite{gu2019tiresias}; (ii) Tiresias (Single)~\cite{gu2019tiresias, hu2023lucid}; (iii) Gavel~\cite{narayanan2020heterogeneity}. Tiresias~\cite{gu2019tiresias} uses 2D-LAS to perform fair sharing of the cluster. Tiresias (Single) employs the Tiresias~\cite{gu2019tiresias} scheduling policy and uses \SK for job packing; however, similar to Lucid~\cite{hu2023lucid} and Pollux~\cite{qiao2021pollux} it does not pack distributed jobs by default. 
Gavel~\cite{narayanan2020heterogeneity} formulates scheduling as an optimization framework that supports the LAS policy and incorporates job packing.

In addition, we also compare \SK with Gavel-FTF~\cite{narayanan2020heterogeneity} to study how \SK affects fairness metrics. Gavel-FTF~\cite{narayanan2020heterogeneity} performs job packing and solves an optimization problem associated with the finish-time fairness (FTF) metric.

\paragraph{Configuration.}
\SK similar to prior works~\cite{zheng2023shockwave, narayanan2020heterogeneity} is a round-based scheduler. We set the round duration to six minutes. \SK is designed as a modular packing policy that can work with any scheduling policy.  We evaluate  \SK by combining it with Tiresias (\ST), and  FTF (\SFair).  \ST denotes using Tiresias scheduling with \SK as a placement policy. \SFair denotes using FTF scheduling with
\SK.

\paragraph{Performance Metrics.}
Similar to the prior DL schedulers, we report 
standard metrics: Average job completion time (Avg. JCT), and the time needed to complete all jobs (Makespan). Additionally, we also evaluate the finish-time fairness (FTF) ratio to capture fairness. 

\subsection{End-to-End Real Cluster Experiments}
\label{sec:real_cluster}

In this section, we evaluate \SK in both cluster and simulation settings. 

\begin{figure}[t]
\centering
\subfigure[Scheduling Efficiency]{
\includegraphics[width=0.43\linewidth]{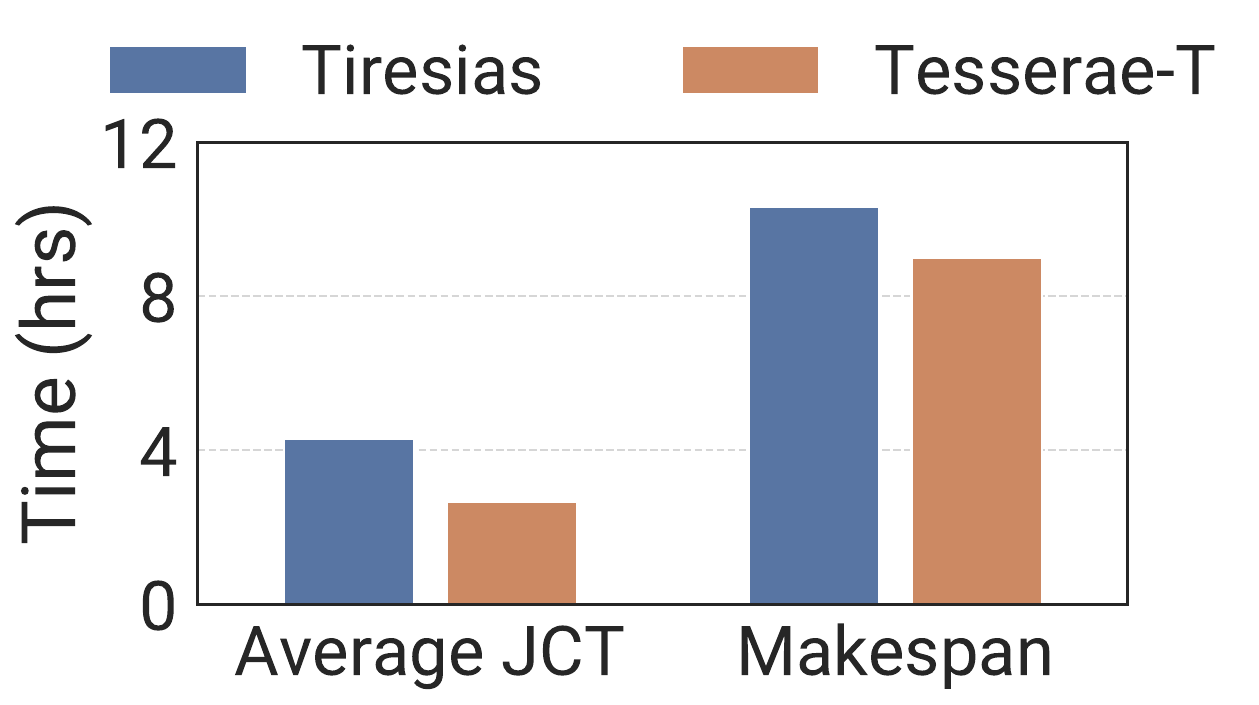}
\label{fig:physical}
}
\subfigure[CDF for Scheduler]{  
\includegraphics[width=0.43\linewidth]{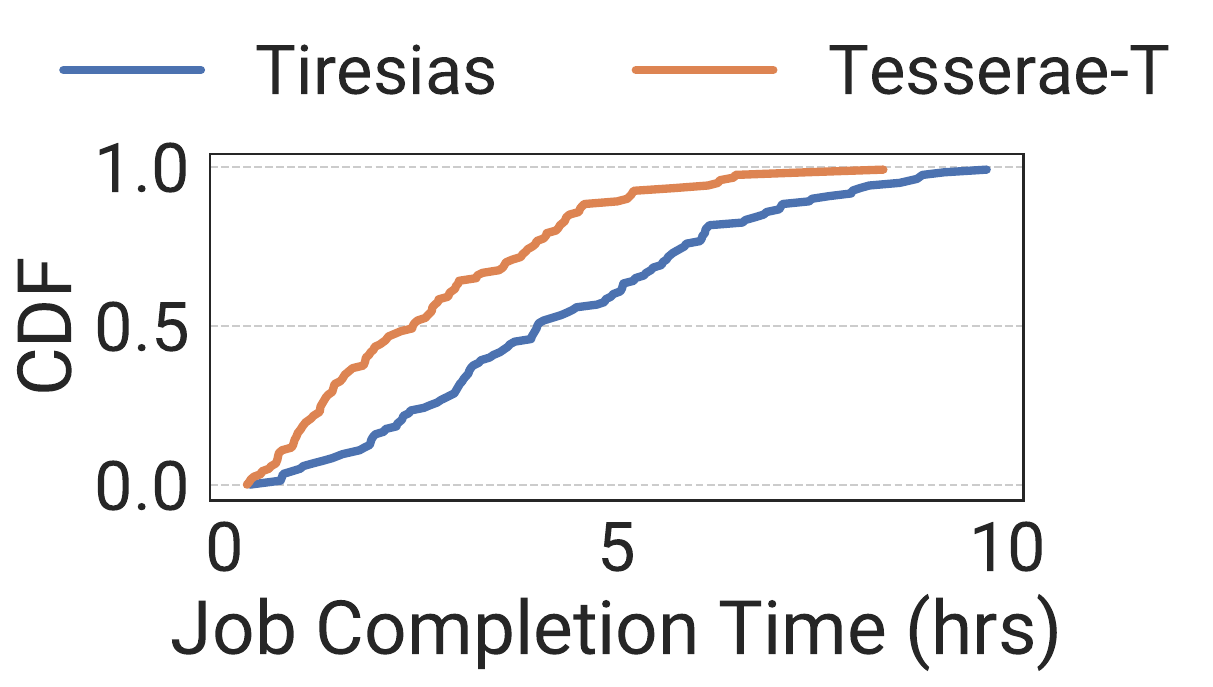} 
\label{fig:physical_cdf}
}
\vspace{-10pt}
\caption{\small{\textbf{Physical cluster evaluation:} We evaluate \ST against \TS on a 32-GPU physical cluster. Compared to \TS, \ST improves Avg. JCT  by $1.62\times$ and Makespan by $1.15\times$.}}
\vspace{-10pt}
\label{fig:physical_bar_cdf}
\end{figure}

\paragraph{\SK Comparison.}
\label{sec:physical_cluster_comparison}
To evaluate the benefits of \SK, we first run \ST and compare it against Tiresias on a 32-GPU physical cluster. In Figure~\ref{fig:physical}, we show that \ST can improve Avg. JCT by $1.62\times$, and Makespan by $1.15\times$ on physical clusters. Figure~\ref{fig:physical_cdf} shows a CDF of JCTs. In Figure~\ref{fig:physical_cdf}, we observe that \ST can significantly reduce the Avg. JCT for jobs with a short duration, which is especially impactful given that Tiresias (and LAS scheduling) is designed to prefer short jobs.

\begin{figure}[t]
\centering
\subfigure[Tiresias]{
\includegraphics[width=0.43\linewidth]{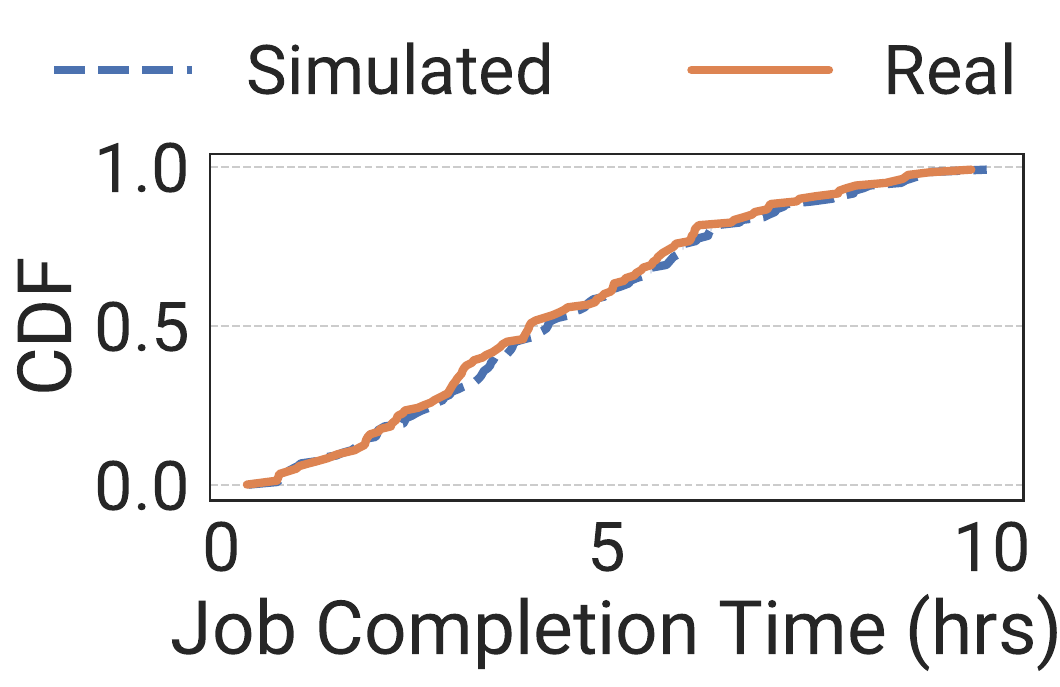}
\label{fig:tiresias_cdf}
}
\subfigure[\ST]{  
\includegraphics[width=0.43\linewidth]{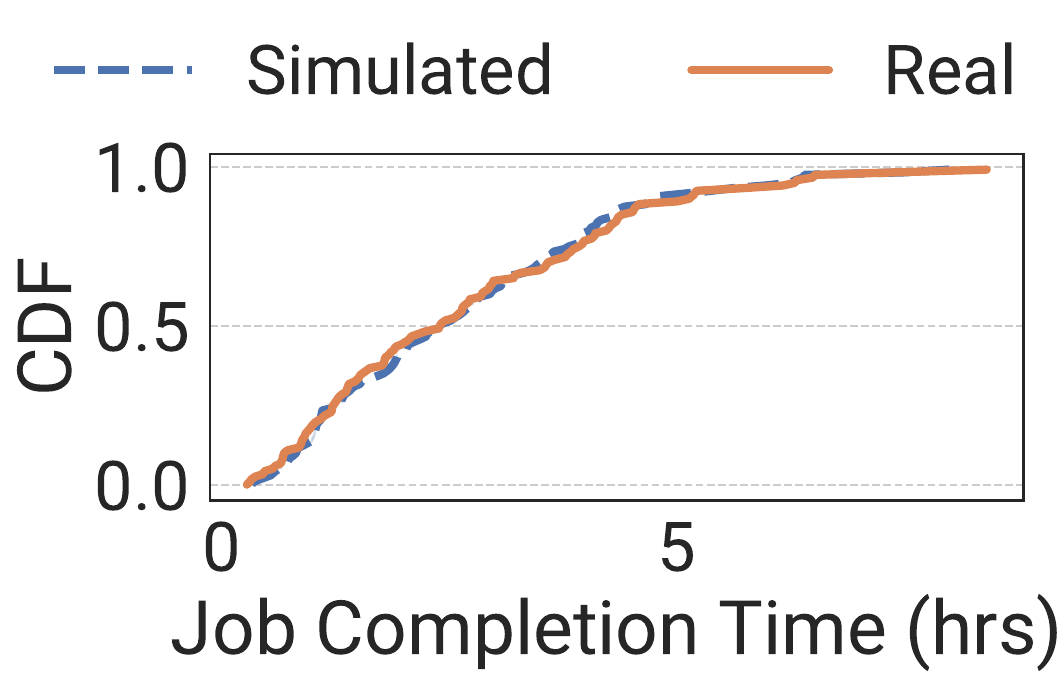} 
\label{fig:matchmaven_tiresias_cdf}
}
\vspace{-10pt}
\caption{\small{\textbf{Comparison of CDFs between cluster and simulator:} We depict the CDF for Tiresias and \ST obtained from physical experiments compared with simulated results. The results demonstrate our simulator's low fidelity.}}
\vspace{-10pt}
\label{fig:physical_simulation_cdf}
\end{figure}

\paragraph{Simulating \SK.} 
Due to lack of access to large scale production clusters, prior works~\cite{qiao2021pollux, narayanan2020heterogeneity, jayaram2023sia, zheng2023shockwave, hu2023lucid, zhao2022multi, gu2019tiresias, mahajan2020themis, mohan2022looking, weng2022mlaas} have used simulations to perform detailed study of large scale traces and compare different metrics. We follow a similar approach.

First, we verify that our simulation closely approximates runs on a real cluster. Since profiling can often have significant noise when performing packing, we run profiling five different times and during simulation we choose one of them at random.
To account for noise, we also run the simulation five different times. In Table~\ref{tab:physical}, we show that our simulation shows the maximum average deviation for JCT between physical cluster and simulation is 3.36\% and the maximum deviation for makespan is 5.42\%. 

In Figure~\ref{fig:physical_simulation_cdf}, we also randomly sample one simulation run and present the CDF of JCTs to highlight the accuracy of our simulation. The average JCT deviation between the physical cluster and simulation results for \ST is 0.21\%, as shown in Figure~\ref{fig:physical_simulation_cdf}.

\begin{table}[t]
\centering
\caption{\small{\textbf{The Fidelity of simulator:}  We run the simulation five different times and depict the mean deviation and standard deviation. We observe the maximum deviation being 5.42\% highlighting that our simulator closely follows the real cluster. }}
\vspace{-10pt}
{
\begin{tabular}{c c c c}
\toprule
Method & Avg. JCT (s) & Makespan (s) \\ 
\midrule
\TS & 3.36\% $\pm$ 0.46\% & 2.05\% $\pm$ 0.03\% \\
\ST & 0.35\% $\pm$ 0.33\% & 5.42\% $\pm$ 0.95\% \\
\bottomrule
\end{tabular}
}
\vspace{-10pt}
\label{tab:physical}
\end{table}

\subsection{End-to-End Results in Simulation}
\label{sec:simulation}
To further evaluate \SK, we use simulation on a large 900 job trace with an 80 GPU cluster. We first evaluate the effect of the migration and packing algorithms introduced in \S\ref{sec:migration} and \S\ref{sec:optimization}  respectively to isolate the performance benefits of \SK. Next, we investigate the adaptability and compatibility of \SK with various hardware and schedulers. Finally, we evaluate the scalability of \SK by varying the number of active jobs and analyzing the overhead introduced by each policy.

\begin{figure}[!t]
    \centering
    \includegraphics[width=0.8\linewidth]{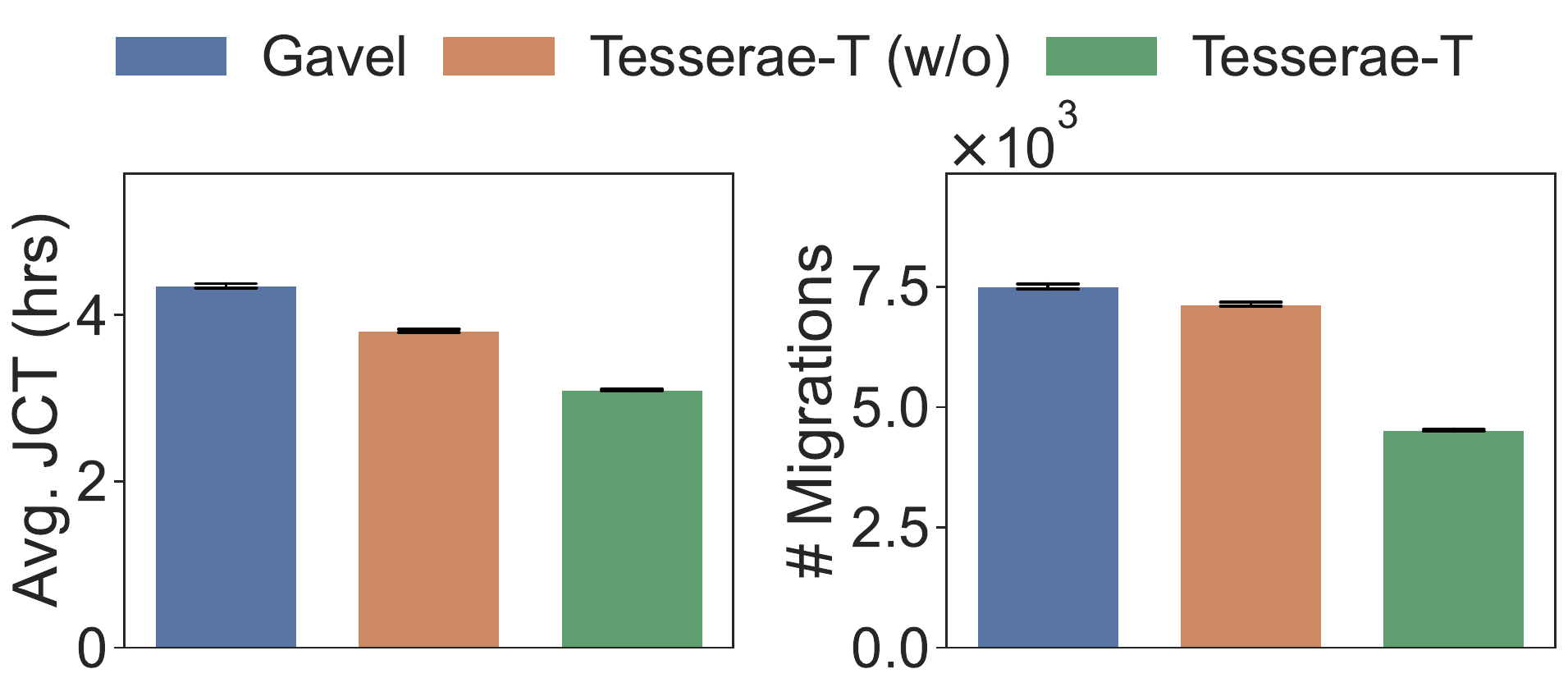}
    \vspace{-10pt}
    \caption{\small{\textbf{Evaluating \ST against optimization-based solutions:} We use w/o to denote the use of the basic migration algorithm described in~\cite{narayanan2020heterogeneity}. First, we notice that our packing policy and migration policy improve the Avg. JCT by $1.41\times$ for \ST compared with Gavel. Second, we observe that our migration policy reduces the number of migrations by 36\% for \ST.}}
    \vspace{-10pt}
    \label{fig:performance_opt}
\end{figure}

\paragraph{Performance Comparison against Optimization Solutions.} Figure~\ref{fig:performance_opt} shows that our packing policy improves the Avg. JCT by $1.15\times$ compared to the optimization-based scheduler Gavel~\cite{narayanan2020heterogeneity}. \ST leverages Tiresias as the scheduling policy while aiming to use our graph-based packing policy (\S\ref{sec:optimization}) to 
maximize total throughput. In addition, we also observe that our migration algorithm outlined in~\S\ref{sec:migration} reduces the migrations of \SK by 36\%, compared to the basic migration algorithm used in~\cite{narayanan2020heterogeneity}. Furthermore, the reduced migration improves  Avg. JCT by $1.22\times$. This highlights that reducing migrations can significantly enhance scheduling efficiency. This also indicates that schedulers should also account for minimizing migrations.


\begin{figure}[t]
\centering
\subfigure[NERSC A100 Machine]{
\includegraphics[width=0.43\linewidth]{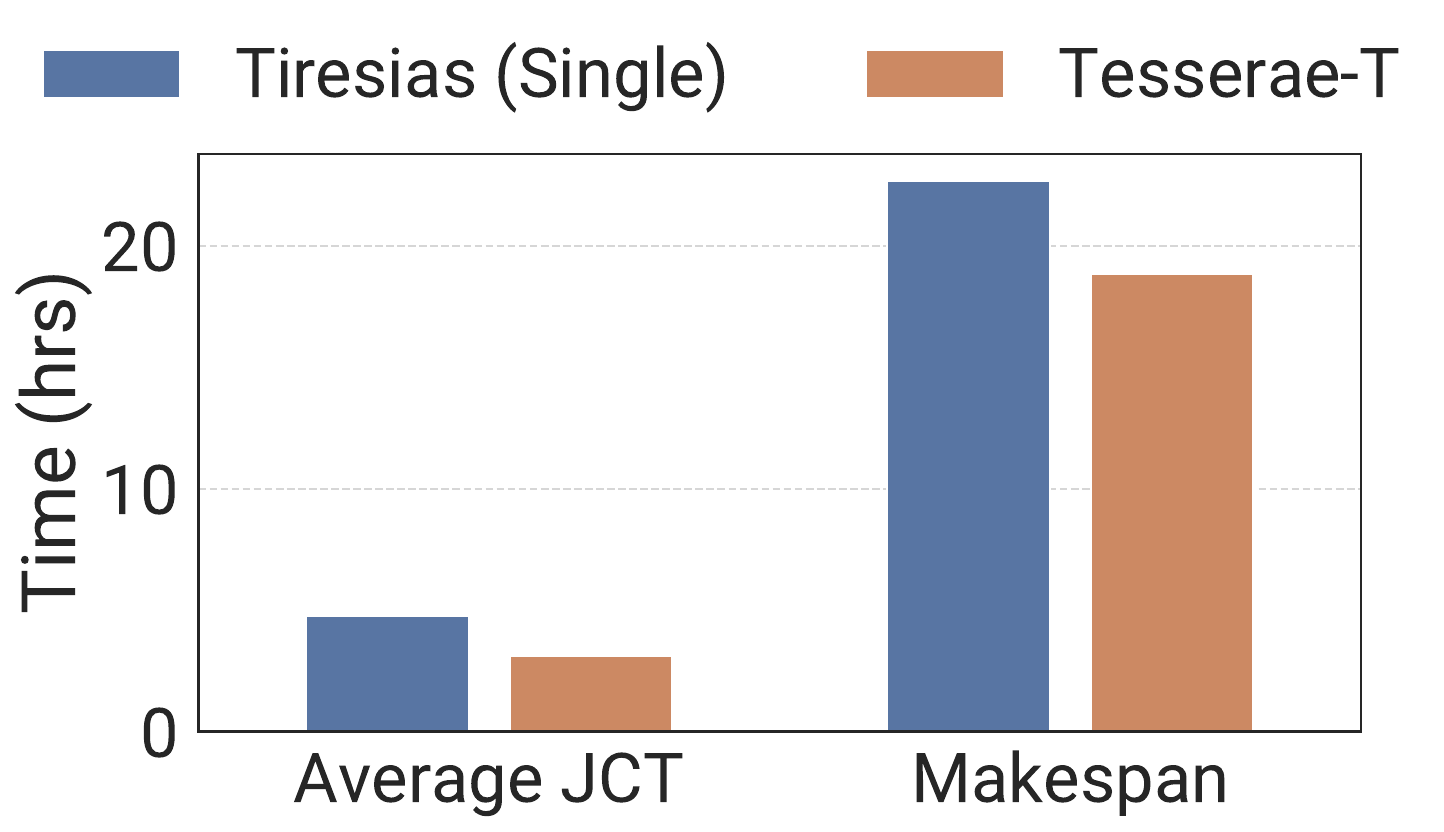}
\label{fig:heuristic_A100}
}
\subfigure[AWS V100 Machine]{  
\includegraphics[width=0.43\linewidth]{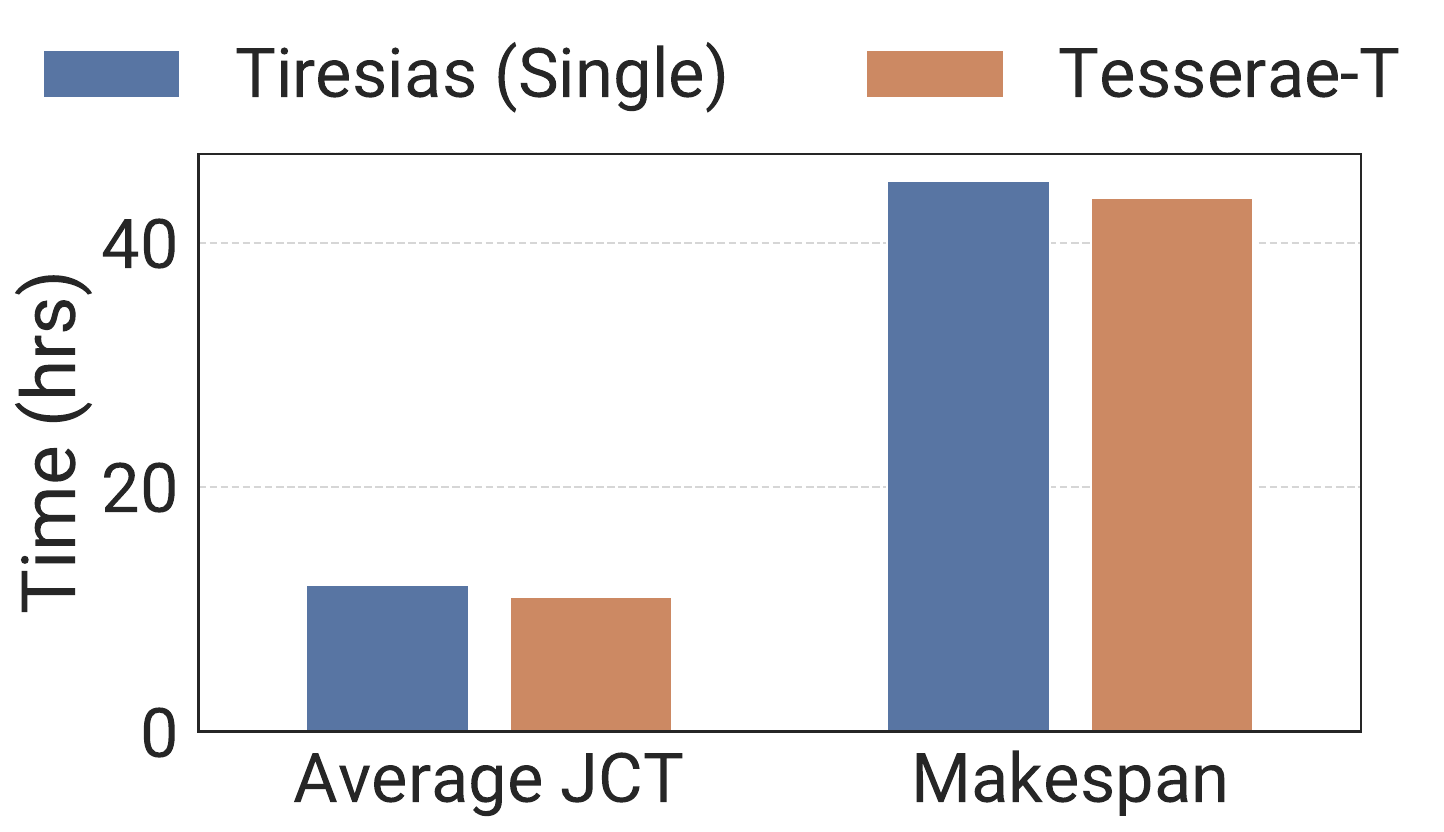} 
\label{fig:heuristic_V100}
}
\vspace{-10pt}
\caption{\small{\textbf{Evaluating \SK against heuristic solution:} Tiresias (Single) employs the Tiresias scheduling policy~\cite{gu2019tiresias} and utilizes \SK for job packing; however, following~\cite{hu2023lucid}, it defaults to packing only 1-GPU jobs. Experimental results demonstrate that \SK improves the Avg. JCT and makespan by up to $1.54\times$ and $1.20\times$, respectively, compared to Tiresias (Single).}}
\vspace{-10pt}
\label{fig:heuristic}
\end{figure}

\paragraph{Performance Comparison against Heuristic Methods} Figure~\ref{fig:heuristic} compares \SK with Tiresias (Single), which uses Tiresias for scheduling and applies the packing policy from \S\ref{sec:optimization} only to 1-GPU jobs due to network contention. Figure~\ref{fig:heuristic_A100} shows that \SK improves Avg. JCT by $1.54\times$ and reduces makespan by $1.20\times$ compared to Tiresias (Single), by leveraging more packing opportunities to increase per-round throughput. This highlights the effectiveness of our packing policy in enhancing scheduling efficiency.
\paragraph{Adaptability of \SK.} To evaluate \SK can adapt to changing hardware we switch our experiment testbed to V100 GPUs using AWS p3.16xlarge instances.
We use the same workload used in Figure~\ref{fig:heuristic_A100}. We observe that \SK can easily adapt to changing hardware and as shown in Figure~\ref{fig:heuristic_V100} improves the Avg. JCT and Makespan by $1.08\times$ and $1.03\times$ respectively.
Compared to the results on the A100 GPU, the V100 GPU’s lower performance and limited memory capacity reduce packing opportunities, thereby diminishing the overall scheduling gains.  \SK adapts to this changing hardware and outperforms heuristic methods.

\begin{figure}[t]
    \centering
    \includegraphics[width=0.7\linewidth]{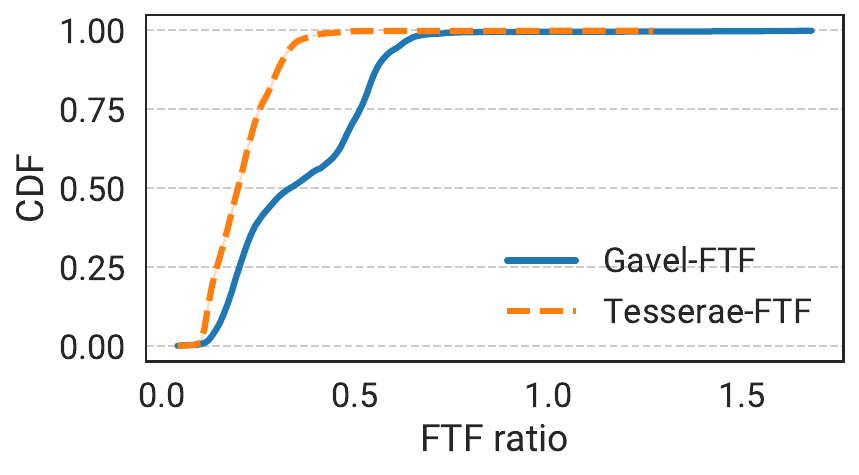}
    \vspace{-10pt}
    \caption{\small{\textbf{Evaluation \SFair's fairness:} The CDF of Finish-time fairness (FTF) ratio~\cite{mahajan2020themis}. The results indicate that \SFair achieves the lowest worst-case FTF ratio, outperforming Gavel-FTF.}}
    \vspace{-10pt}
    \label{fig:fairness}
\end{figure}

\paragraph{Compatibility with Other Schedulers} \SK is implemented as a modular packing and placement plugin on top of existing schedulers. We can use \SK over existing schedulers without modifying the underlying scheduler.
To show modularity of \SK and it's impact on underlying metric, we implement \SK over Gavel-FTF. To evaluate fairness, similar to prior work~\cite{mahajan2020themis} we use FTF ratio as a metric. FTF is defined as $\rho = \frac{T_{s}}{T_{f}}$, where $T_{s}$ is the job completion time in a shared cluster and $T_{f}$ is the job completion time in an isolated and fairly shared cluster. In Figure~\ref{fig:fairness},
we show that \SFair can enhance the performance of Gavel-FTF. This highlights that \SFair also provides higher fairness than existing baselines. 

\begin{figure}[!t]
\centering
\subfigure[Overhead]{
\includegraphics[width=0.43\linewidth]{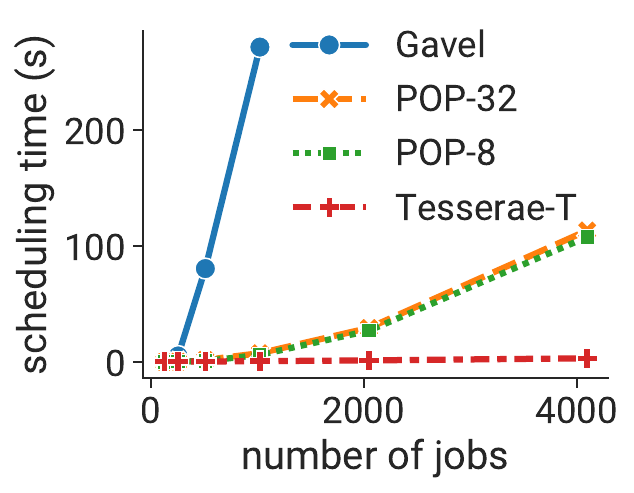}
\label{fig:scheduler_overhead}
}
\subfigure[Breakdown]{  
\includegraphics[width=0.43\linewidth]{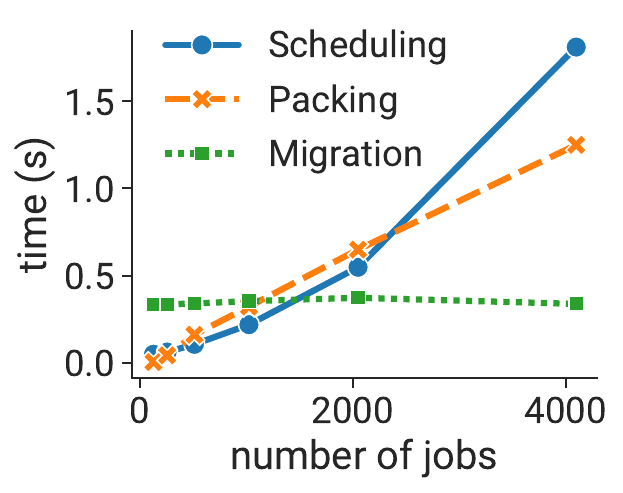} 
\label{fig:scheduler_breakdown}
}
\vspace{-10pt}
\caption{\small{\textbf{Scalability of Schedulers:} The left figure shows the overhead of \ST compared with Gavel~\cite{narayanan2020heterogeneity} and POP~\cite{narayanan2021solving} with the increased number of active jobs. The right figure presents the overhead breakdown of \ST.}}
\vspace{-10pt}
\label{fig:overhead_exp}
\end{figure}

\paragraph{Scalability Analysis} We fix the number of GPUs in the cluster and vary the number of active jobs to evaluate the scalability of \SK. In Figure~\ref{fig:scheduler_overhead}, we observe that \SK scales better than Gavel~\cite{narayanan2020heterogeneity} and POP~\cite{narayanan2021solving} as the number of active jobs increases significantly. Moreover, Figure~\ref{fig:scheduler_breakdown} presents a breakdown of \ST's overhead. We notice that the overhead of scheduling and packing increases with the number of active jobs, while the migration overhead remains stable. This is because the scheduling and packing algorithms scale with the number of active jobs, whereas the migration overhead depends on the number of GPUs in the cluster.


\section{Ablation Studies}
\label{sec:ablation}

In this section, we investigate various parameters of \SK and evaluate the packing and parallelism strategy with ablation experiments.

\subsection{Impact of Parallelization Strategy} 
\label{sec:parallelism}
\begin{figure}[!t]  
    \centering
    \includegraphics[width=0.8\linewidth]{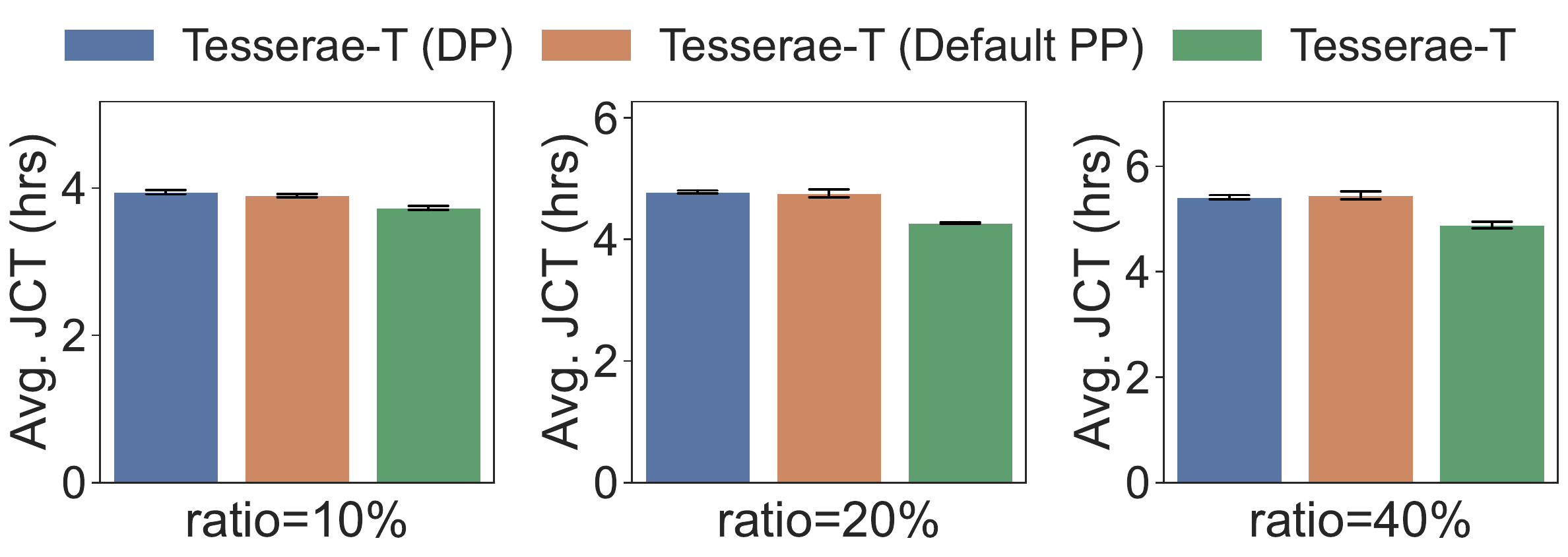}
    \vspace{-10pt}
    \caption{
    \small{\textbf{Impact of Parallelization strategy:} We compare the impact of parallelism strategy on Avg. JCT of Large Language Models (GPT3-Medium, GPT3-XL, and GPT3-3B). The Default PP (Def PP) is provided by Megatron-LM~\cite{shoeybi2019megatron}. \ST selects the best parallelism strategy from DP, TP, and the candidate of possible PP strategies. By varying the ratio of large language models in the workload, we observe that selecting the best parallelism strategy can improve Avg. JCT of large language models by $1.12\times$.}}
    \vspace{-10pt}
    \label{fig:frac_parallelism}
\end{figure}

To highlight the impact parallelization strategies can have on the throughput of packed jobs, we compare \ST (DP), \ST (Default PP), and \ST in this experiment. \ST (DP) selects data parallelism as the parallelism strategy for packed language model training jobs. The parallelism strategy selected by \ST (Default PP) is the default pipeline parallelism strategy used in Megatron-LM~\cite{shoeybi2019megatron}. In contrast, \ST picks the best parallelism strategy from the candidate set defined by users. As demonstrated in Figure~\ref{fig:frac_parallelism}, by adjusting the parallelism strategy for packed jobs, there is a 12\% improvement in Avg. JCT for large language models.

\subsection{Parameter Sensitivity}
\label{sec:parameter}

\begin{figure}[t]
    \centering
    \includegraphics[width=0.8\linewidth]{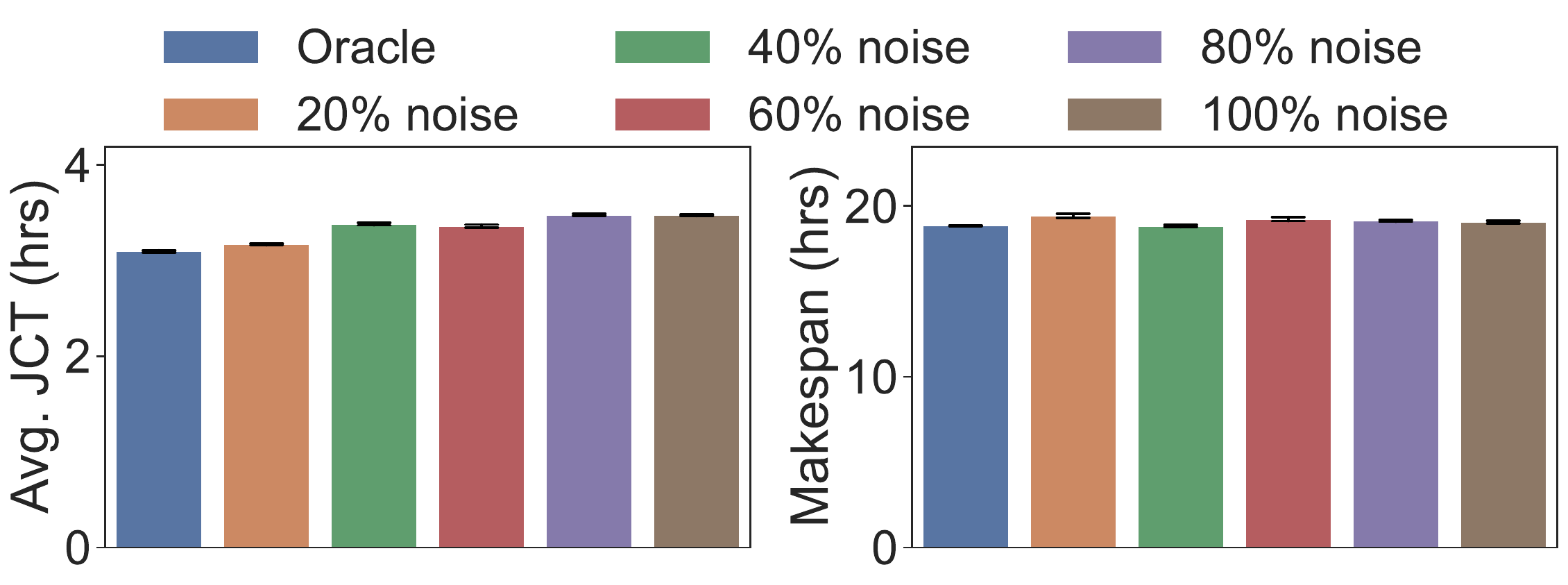}
    \vspace{-10pt}
    \caption{\small{\textbf{Impact of inaccurate profiling on \ST:} Our results indicate that \ST is robust to noise in profiling data, even when the noise is 100\%. }}
    \label{fig:tiresias_noise}
    \vspace{-10pt}
\end{figure}

\paragraph{Sensitivity to Profiling Errors.} It is possible that the profiling results are not correct due to software or hardware variabilities~\cite{maricq2018taming,schad2010runtime}. In this experiment, the profiling data we used to make packing decisions is multiplied by a random factor sampled from $[1-n_{p}, 1+n_{p}]$, where $n_{p}$ is a noise parameter from $[0, 1]$. From the results shown in Figure~\ref{fig:tiresias_noise}, we notice that the Avg. JCT is increased by at most $1.12\times$ with 100\% noise, while the Makespan is robust with the increased noise parameter $n_{p}$. However, in reality, we find the profiling noise is always under 20\%,
and thus the profiling noise has little impact on \SK.

\begin{figure}[t]
\centering
\includegraphics[width=0.8\linewidth]{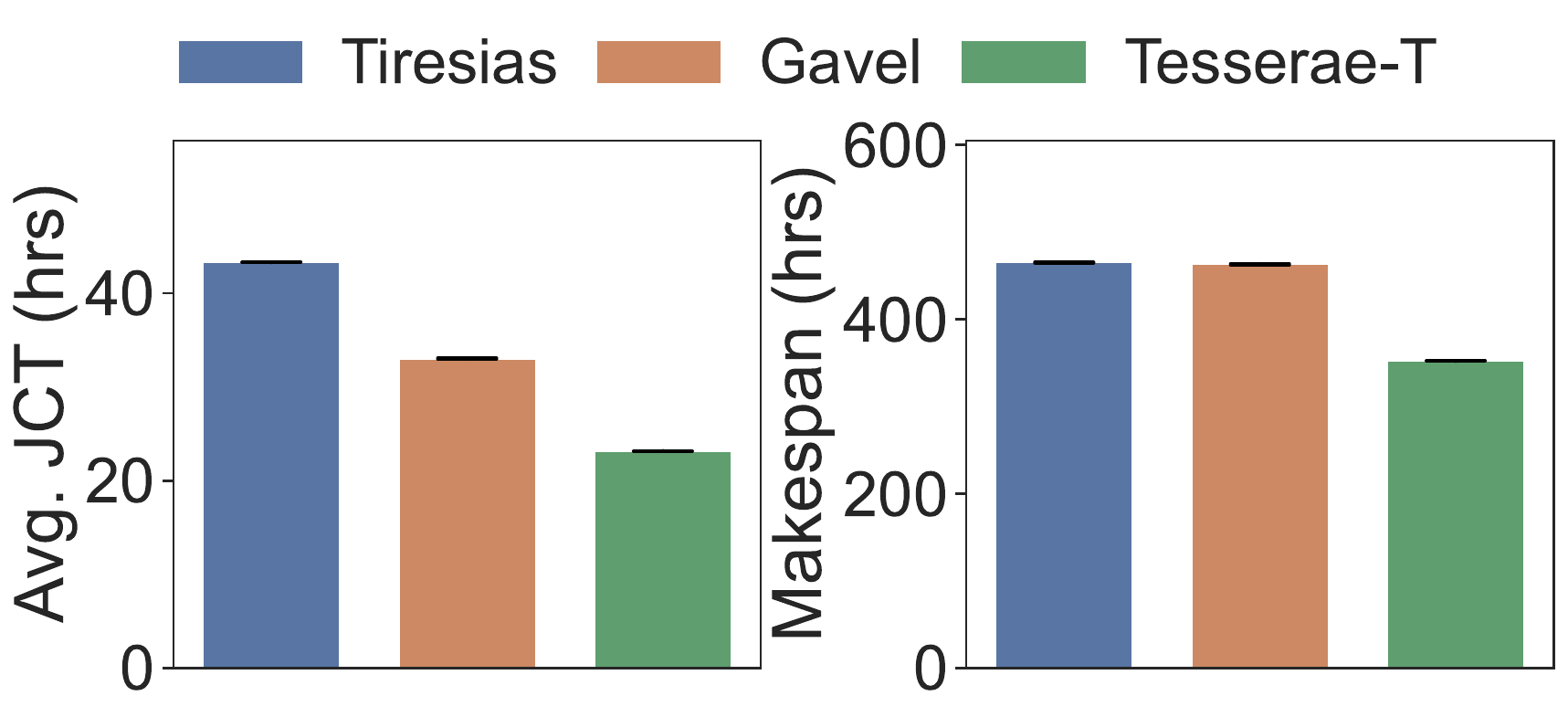}
\vspace{-10pt}
\caption{\small{\textbf{Evaluating \ST's scheduling efficiency by varying the workload:} We evaluate \ST's scheduling efficiency on a large-scale cluster with 80 GPUs over trace generated by Gavel's trace generator. This trace holds 900 jobs but follows a different duration distribution. The results show that \ST improves Avg. JCT by up to $1.87\times$ compared with existing scheduling algorithms.}}
\vspace{-10pt}
\label{fig:tiresias_simulation_gavel_workload}
\end{figure}

\paragraph{Sensitivity to Workload.} We also compare \ST with other baselines on a 900 jobs trace, which is generated by Gavel's trace generator, with an 80-GPU cluster. For this trace, pursuant to Gavel, the duration of jobs is uniformly sampled between $10^{[1.5, 3]}$ minutes with 80\% probability, and the remaining 20\% jobs have their duration uniformly sampled $10^{[3, 4]}$ minutes. Similar to Gavel trace, 70\% of the jobs request a single GPU, 10\% of the jobs request 2 GPUs, 15\% of the jobs request 4 GPUs, and the remaining 5\% of the jobs request 8 GPUs. From Figure~\ref{fig:tiresias_simulation_gavel_workload}, we have a similar observation, \ST outperforms all evaluated baselines over performance metrics. Specifically, \ST can reduce Avg. JCT, and Makespan by up to $1.87\times$, and $1.32\times$ respectively.

\begin{figure}[t]
    \centering
    \includegraphics[width=0.8\linewidth]{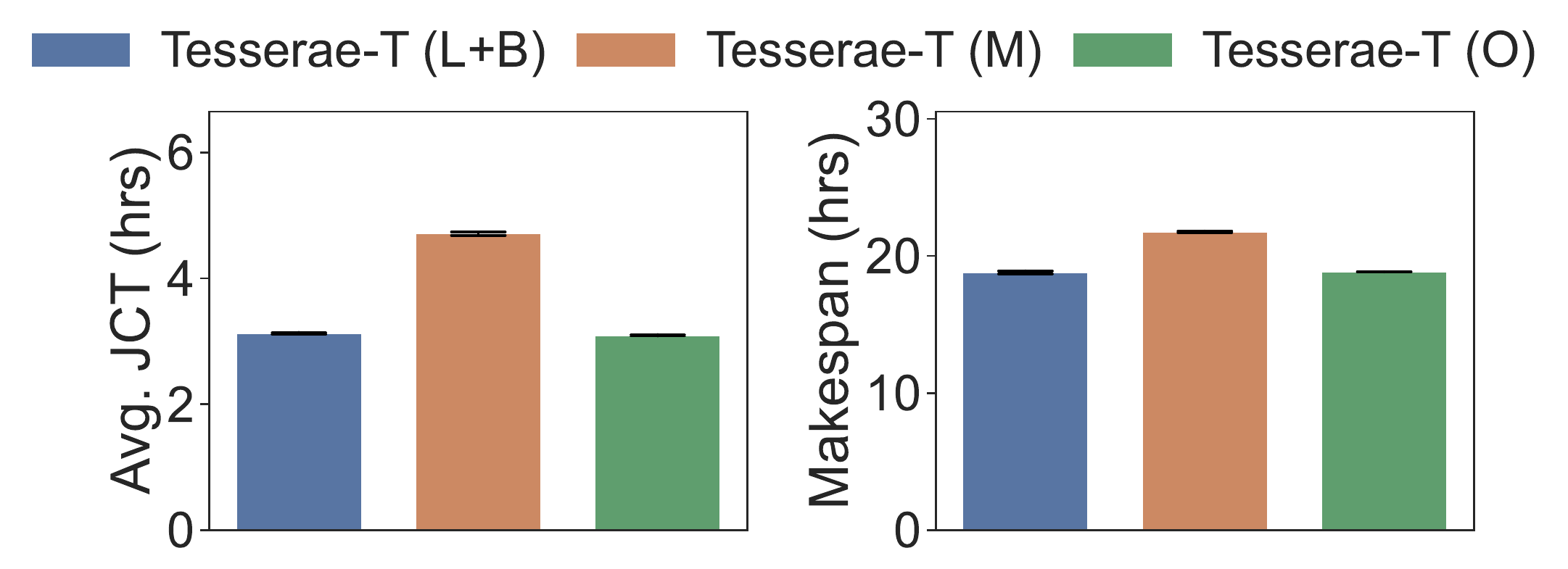}
    \vspace{-10pt}
    \caption{
    \small{\textbf{Reduce Profiling Cost:} We compare our throughput estimator (\underline{L}inear model and \underline{B}ayesian optimization) with \underline{M}atrix Completion and \underline{O}racle. The results show that our throughput estimator can be used to reduce profiling costs and maintain scheduling efficiency.}}
    \vspace{-10pt}
    \label{fig:profile_cost}
\end{figure}

\paragraph{Reduce Profiling Cost.} It is expensive to profile each model and each possible model combination. To reduce the profiling cost, we utilize a linear model~\cite{jayaram2023sia} to estimate the throughput for the data parallel applications and bayesian optimization~\cite{snoek2012practical} to predict the throughput of LLM workloads with varying parallelism strategies. Figure~\ref{fig:profile_cost} shows that our strategy outperforms the matrix completion method used in Gavel~\cite{narayanan2020heterogeneity} and Quasar~\cite{delimitrou2014quasar}. In addition, our throughput estimator can predict missing throughput with only a minor reduction in Avg. JCT compared to Oracle, which involves offline profiling of each model and each model combination.

\section{Related Work}
\label{sec:related}

\paragraph{GPU Cluster Schedulers.} GPU cluster schedulers have been actively researched in recent years~\cite{peng2018optimus, xiao2020antman, hwang2021elastic, chaudhary2020balancing, hu2023lucid, jayaram2023sia, li2023lyra, li2023easyscale, choudhury2024mast}. In detail, Gandiva~\cite{xiao2018gandiva} uses time and space sharing to improve cluster utilization. Tiresias~\cite{gu2019tiresias} utilizes two-dimensional indexes to reduce queuing delays. Themis~\cite{mahajan2020themis} and Shockwave~\cite{zheng2023shockwave} design algorithms to trade off fairness for efficiency. Pollox~\cite{qiao2021pollux} co-optimizes system throughput and statistical efficiency to minimize average JCT. Synergy~\cite{mohan2022looking} allocates CPU and memory to sensitive Jobs. Muri~\cite{zhao2022multi} considers multi-resource interleaving to improve resource utilization. The closest scheduler to us is Gavel~\cite{narayanan2020heterogeneity}. In contrast, Gavel~\cite{narayanan2020heterogeneity} formalizes each policy as an optimization problem and does not consider optimizing the parallelism strategy for foundation model training with space sharing.

\paragraph{Resource Sharing for DL Workloads.} GPU sharing is one possible way to improve GPU utilization, which has been deployed in GPU cluster schedulers~\cite{xiao2018gandiva, weng2022mlaas, xiao2020antman, narayanan2020heterogeneity}. Here, we introduce several GPU-sharing tools. Multi-Process Service~\cite{mps} and Multi-Instance GPU~\cite{mig} enable multiplex jobs on NVIDIA GPUs. PipeSwitch~\cite{bai2020pipeswitch} and Salus~\cite{yu2020fine} are designed for fast job switching and memory sharing. Zico~\cite{lim2021zico} efficiently shares memory among packing jobs without exceeding a given memory budget. MuxFlow~\cite{zhao2023muxflow} supports safe and fast space-sharing in the production cluster. 

\paragraph{Foundation Models.} Foundation models, first introduced by~\cite{vaswani2017attention}, are effective in various language understanding tasks such as text generation, text classification, and question answering~\cite{wang2018glue, rajpurkar2016squad}. Recent works, e.g. GPT~\cite{radford2019language, brown2020language}, LLaMA~\cite{touvron2023llama, dubey2024llama, touvron2023llama2}, Qwen~\cite{bai2023qwen, yang2024qwen2}, and Gemma~\cite{team2024gemma, team2024gemma2} have shown that scaling foundation models can achieve high accuracy on many downstream tasks. 

\paragraph{Parallelism for Foundation Models.} Due to the increased size of foundation models and the limited GPU memory, various model parallelism techniques have been proposed to train foundation models~\cite{narayanan2021efficient}. These techniques can be divided into two major classes: tensor model parallelism~\cite{jia2019beyond, shazeer2018mesh} and pipeline model parallelism~\cite{huang2019gpipe, narayanan2019pipedream, narayanan2021memory}. The application of these methods has facilitated the development of Megatron-LM~\cite{shoeybi2019megatron} and DeepSpeed~\cite{rasley2020deepspeed}, contributing to the acceleration of distributed training in foundation models. Recently, Alpa~\cite{zheng2022alpa} has been developed for searching optimal execution plans for distributed training. However, this paper studies distributed training with space sharing in GPU cluster scheduling. 

\section{Conclusion}
\label{sec:conclusion}

In this paper, we propose \SK, a general GPU cluster scheduler that supports various existing scheduling policies. Rather than formulating the packing problem as an optimization problem, we develop an innovative and efficient packing algorithm inspired by the Hungarian algorithm. Furthermore, \SK is the first scheduler to facilitate the packing of 3D Parallelism training jobs. We also explore opportunities for optimizing the parallelism strategy for packing jobs. Lastly, \SK develops a novel migration algorithm to reduce migrations during scheduling. Our physical and simulated experiments show that \SK outperforms existing state-of-the-art schedulers.

\bibliographystyle{plain}
\bibliography{sample-base}






\clearpage
\appendix
\section{Examples of Migration Algorithm}
\label{sec:appendix_example}

In this section, we present several examples to demonstrate the principles of the Algorithm~\ref{alg:job-migration} and Algorithm~\ref{alg:node-level}.

\begin{example}
Consider two placement plans from consecutive rounds $i$ and $i+1$ as follows: $\{(0,1), (1,2), (2,3), (3,4)\}$ and $\{(0,4), (1,1), (2,2), (3,3)\}$. In each tuple, the first number represents the GPU ID, and the second number represents the job ID. Since Jobs $1$-$4$ appear in two placement plans, we do not need to remove any jobs. Then, we compute the migration cost matrix by using Algorithm~\ref{alg:node-level}:
\begin{align*}
    C_{\text{node}} = \begin{bmatrix}
    1 & \underline{0} & 1 & 1 \\
    1 & 1 & \underline{0} & 1 \\
    1 & 1 & 1 & \underline{0} \\
    \underline{0} & 1 & 1 & 1 \\
    \end{bmatrix}
\end{align*}
For example, the value of $C[0, 2]$ is $1$ because if we reassign GPU $0$ to $2$, we must move job $1$ out and move job $2$ in. The cost for move-in and move-out for the 1-GPU job is $0.5$ respectively. Therefore, the total cost is $0.5 + 0.5 =1$. Using the Hungarian algorithm, we get that the minimized migration is $0$.
\end{example}

\begin{example}
Consider two placement plans from consecutive rounds $i$ and $i+1$ as follows: $\{(0,(1,5)), (1,2), (2,3), (3,4)\}$ and $\{(0,(4,5)), (1,1), (2,2), (3,3)\}$. In each tuple, the first number represents the GPU ID, and the second number/tuple represents jobs on the GPU. For instance, $(1,5)$ means Job $1$ and Job $5$ are placed on the GPU $0$. Then, we compute the migration cost matrix by using Algorithm~\ref{alg:node-level}:
\begin{align*}
    C_{\text{node}} = \begin{bmatrix}
    1 & \underline{0.5} & 1.5 & 1.5 \\
    1.5 & 1 & \underline{0} & 1 \\
    1.5 & 1 & 1 & \underline{0} \\
    \underline{0.5} & 1 & 1 & 1 \\
    \end{bmatrix}
\end{align*}
For example, the value of $C[0, 1]$ is $0.5$ because if we reassign GPU $0$ to $1$, we must move job $5$ out. The cost for move-in and move-out for the 1-GPU job is $0.5$ respectively. Therefore, the total cost is $0.5$. Using the Hungarian algorithm, we get that the minimized migration is $1$. This indicates that job $5$ must be relocated from being colocated with job $1$ to being colocated with job $4$.
\end{example}

\begin{example}
Consider two placement plans from consecutive rounds $i$ and $i+1$ as follows: $\{(0,(1,6)), (1,2), (2,3), (3,4)\}$ and $\{(0,(4,5)), (1,1), (2,2), (3,3)\}$. In each tuple, the first number represents the GPU ID, and the second number/tuple represents jobs on the GPU. Since Job $5$ and Job $6$ do not appear in two rounds concurrently, we remove them from the placement plan first. Then, we compute the migration cost matrix by using Algorithm~\ref{alg:node-level}:
\begin{align*}
    C_{\text{node}} = \begin{bmatrix}
    1 & \underline{0} & 1 & 1 \\
    1 & 1 & \underline{0} & 1 \\
    1 & 1 & 1 & \underline{0} \\
    \underline{0} & 1 & 1 & 1 \\
    \end{bmatrix}
\end{align*}
Using the Hungarian algorithm, we get that the minimized migration is $0$.
\end{example}


\section{Discussion of Migration Algorithm}
\label{sec:discuss_migration}

We also have a simpler migration algorithm for non-packing placement plans, which is presented in Algorithm~\ref{alg:job-migration-non-packing}.

\begin{algorithm}[!ht]
    \caption{\textsf{Job Migration for Non-packing Placement Plans}}
    \label{alg:job-migration-non-packing}
    \KwIn{\textsf{placement\_plan for round $i$: $P_i$}, \textsf{placement\_plan for round $i+1$: $P_{i+1}$}.}
    \KwOut{\textsf{Migration Plan} \textsf{M}}
    $C \leftarrow [0]_{k \times k}$, where $k$ is the number of GPUs in the node. \\
    \textsf{Remove all jobs $j$ in $P_i$ and $P_{i+1}$, where job j satisfies} $j \in ((P_i \cup P_{i+1}) - (P_{i} \cap P_{i+1}))$ \\    
    \ForEach{GPU \textsf{$u$} $\in$ \textsf{$P_{i}$}}{
        \ForEach{GPU \textsf{$v$} $\in$ \textsf{$P_{i+1}$}}{
            $JS_{u} \leftarrow$ Job sets on GPU $u$ and $JS_{v} \leftarrow$ Job sets on GPU $v$ \\
            \ForEach{job \textsf{$j$} $\in JS_u \cup JS_v$}{
                \If{job \textsf{$j$} $\in$ $((JS_u \cup JS_v) - (JS_u \cap JS_v))$}{
                    $C_{u,v} \leftarrow C_{u,v} + 1 / (2 \cdot \textsf{num\_GPUs($j$)})$ \\
                }
            }
        }
    }
    \textsf{$M$} $\leftarrow$ \textsf{Hungarian Algorithm}(\textsf{C}) \\
    \Return \textsf{$M$}
\end{algorithm}

Following the time complexity analysis in \S\ref{sec:shaka}, the time complexity of Algorithm~\ref{alg:job-migration-non-packing} is $\mathcal{O}(k^3)$, where $k$ is the number of GPUs in the cluster. However, the algorithm may violate the consolidated placement settings for packing placement plans. Here, we present an example to illustrate this point.
\begin{example}
Consider two placement plans from consecutive rounds $i$ and $i+1$ as follows: $\{0: \{(0,1), (1,1), (2,1), (3,1)\}, 1: \{ (0,2), (1,2), (2,2), (3,2)\}\}$ and $\{0: \{(0,(1,2)), (1,(1,2)), \\ (2,(1,2)), (3,(1,2))\}\}$. By using Algorithm~\ref{alg:job-migration-non-packing}, we get the following migration cost matrix
\begin{align*}
    \begin{bmatrix}
    0.25 & 0.25 & 0.25 & 0.25 & 0.25 & 0.25 & 0.25 & \underline{0.25} \\
    0.25 & 0.25 & 0.25 & 0.25 & 0.25 & 0.25 & \underline{0.25} & 0.25 \\
    0.25 & 0.25 & \underline{0.25} & 0.25 & 0.25 & 0.25 & 0.25 & 0.25 \\
    0.25 & 0.25 & 0.25 & \underline{0.25} & 0.25 & 0.25 & 0.25 & 0.25 \\
    0.25 & 0.25 & 0.25 & 0.25 & \underline{0.25} & 0.25 & 0.25 & 0.25 \\
    0.25 & 0.25 & 0.25 & 0.25 & 0.25 & \underline{0.25} & 0.25 & 0.25 \\
    0.25 & \underline{0.25} & 0.25 & 0.25 & 0.25 & 0.25 & 0.25 & 0.25 \\
    \underline{0.25} & 0.25 & 0.25 & 0.25 & 0.25 & 0.25 & 0.25 & 0.25 \\
    \end{bmatrix}
\end{align*}
Applying the Hungarian algorithm may yield multiple solutions. However, the solution marked with an underline results in a non-consolidated placement.
\end{example}
However, using Algorithm~\ref{alg:node-level}, we can avoid such cases.

\end{document}